\documentclass[superscriptaddress,twocolumn,prl,showpacs,floatfix]{revtex4}
\usepackage[dvips]{graphicx}
\usepackage{epsfig}
\usepackage{hyperref}
\usepackage{amsmath,amsfonts,amssymb,bm}%,ulem}
\usepackage[caption=false]{subfig}
\begin{document}
\title{Exact solution for low energy quantum anharmonic vibrations in a long polymer chain}
\author{Alexander L. Burin}
\affiliation{Department of Chemistry, Tulane University, New
Orleans, LA 70118, USA}
\date{\today}
\begin{abstract}
We propose the algorithm  for determining quantum stationary states of periodic linear chain of atoms coupled by harmonic and third order anharmonic interactions (Fermi-Ulam-Pasta $\alpha$ problem) in the long wavelength limit within the resonant approach.   These states   can be encoded by  sequences of integer numbers determining their energies and wavefunctions.  Using these states we described the exact  time evolution of a single phonon state showing coherent oscillations. The applications of theory to vibrational energy transport and quantum informatics are discussed. 
\end{abstract}

\maketitle

% cut introduction to half a column. (a) significance due to nanotechnology (b) FPU -> soliton -> resonance appraoch -> quantum problem
Molecular vibrations determine the heat balance in  nano-devices \cite{AbeScience1,AbeScience2} and can be manipulated similarly to electrons and photons and used to carry and process quantum information  \cite{RevModPh,GruebeleQC1,GruebeleQC2,Shapiro}. The molecules composed by chains of self-repeating monomers demonstrate an outstanding ability to transfer  and convert energy because of delocalized normal modes (phonons) existing due to the translational invariance and propagating with the speed of sound as high as $10^{4}$ m$/$s 
%, which can be as high as several thousand meters per second 
in organic polymers \cite{DlottScience,Alkane3,NitzanSegal1,IgorRecent}. 

The increasing theoretical efforts have led to the remarkable progress in understanding polymer  heat conduction (see e. g. 
%Refs. \cite{Turbulence,Leibovitz2008,Shepel2009,PNASRecent} and 
Reviews \cite{RevModPh,DharRev,abPhR,RevRecent}) though the actual mechanisms of energy transport through anharmonic oscillator chains remains unclear since its first numerical study by Fermi, Pasta and Ulam (FPU) \cite{FPUclassic}. The original numerical simulations of vibrational dynamics in the FPU $\alpha$ model including atoms coupled by harmonic and third order anharmonic interactions  were targeted to reveal a thermalization accompanied by the loss of memory about the initial state \cite{FPUclassic}. Instead a quasiperiodic dynamics has been found. This quasiperiodic dynamics has been interpreted introducing the solitary waves solutions  and integrals of motion associated with them \cite{Israilev,RevRecent,Palais,Henrici} suggesting that the system is integrable. 

It is natural to expect that the classical integrability should be reflected in the quantum mechanical properties \cite{Doron}.  Quantum mechanical treatment is important for applications to real molecules since the thermal energy $k_{B}T$ is usually smaller  than the typical vibrational energy which is around $1000$ cm$^{-1}$ \cite{IgorRecent} (even at room temperature $k_{B}T \sim 200$cm$^{-1}$). Here we report the exact quantum mechanical  solution for eigenstates and eigenenergies of the FPU $\alpha$ model in the long-wavelength limit  restricted to  the resonant interactions  \cite{PNASRecent}. Below we introduce the quantum mechanical  model, describe its solution and its application to the single phonon state time evolution. The details of the derivations of the solution are given  in the Supplementary Materials  \cite{SI}.

The normal modes (phonons) of the periodic (circular) chain with the period $a=1$ and length $N$ can be expressed as planar waves with the amplitude depending on the coordinate $z=1, ... N$ along the chain as $x(z)=e^{iqz}/\sqrt{N}$. In a periodic chain one has $x(z+N)=x(z)$ and the wavevector for each phonon can be expressed as $q=2\pi n/N$ with an integer number $n$ %($-N/2<n\leq N/2$, we consider even $N$, while the generalization to odd $N$'s is straightforward) 
identifying each   specific mode. In the long wavelength limit $n \ll N$ the phonon  energy  can be approximated by the linear dispersion law $E_{n}=hc|n|/N$ where $c$ stands for the speed of sound ($c=1$ in the FPU model \cite{SI}). The system Hamiltonian in the harmonic approximation can be conveniently expressed in terms of creation and annihilation operators $b_{n}^{\dagger}, b_{n}$ for each mode $n$  ($\widehat{H}_{0}$ term in Eq. (\ref{eq:H})). 

The third order anharmonic interactions can be introduced using products of three $b$-operators describing the phonon decay into two phonons $b_{m}^{\dagger}b_{n}^{\dagger}b_{m+n}$ or two phonon association backwards $b_{m+n}^{\dagger}b_{m}b_{n}$ conserving the total wavevector  within the long-wavelength (low energy) limit (cf. \cite{PNASRecent,TextPh,SI}). We restrict the consideration to the only fully resonant interactions expressed by the terms with both parameters $m$ and $n$ either positive or negative \cite{PNASRecent}. Indeed, the harmonic energy  does not change in such process  ($h(|m+n|-|m|-|n|)/N=0$) while it changes by a large harmonic energy $2h {\rm min}(|m|, |n|)/N$ in the opposite case. Within this resonant approach the system Hamiltonian can be separated into two parts associated with positive and negative wavevectors. The positive wavevector part can be expressed as \cite{PNASRecent,TextPh} (see also Ref. \cite{SI}, Sec. I; we set the Planck constant $h=1$, the negative wavevector part can be studied similarly)
 \begin{eqnarray}
\widehat{H}_{res} = \widehat{H}_{0}+\frac{\alpha}{\sqrt{2}N^2}\widehat{V}, ~ \widehat{H}_{0}=\frac{1}{N}\sum_{n>0}n(b_{n}^{+}b_{n}+1/2), 
\nonumber\\
\widehat{V}=\frac{1}{2}\sum_{m, n >0}\sqrt{mn(m+n)}\left(b_{m}^{+}b_{n}^{+}b_{m+n} +  b_{m+n}^{+}b_{m}b_{n}\right).
\label{eq:H}
\end{eqnarray}
Here the parameter $\alpha$ is the relative anharmonic interaction used in Ref. \cite{FPUclassic}. %The limits of applicability of our model are defined later in Eq. (\ref{eq:resAppr}). 

Each eigenstate of Eq. (\ref{eq:H}) is determined by a super position of multiphonon states defined by population number sequences $\{\nu\}=(\nu_{1}, \nu_{2}, ... \nu_{n})$ ($\nu_{i}=b_{i}^{\dagger}b_{i}$) as 
\begin{eqnarray}
\psi=\sum_{\{\nu\}} \frac{c(\{\nu\})}{\sqrt{N(\{\nu\})}}|\{\nu\}>, ~
%\nonumber\\
N(\{\nu\})=\prod_{i=1}^{n}i^{\nu_{i}}\nu_{i}!,
\label{eq:EigFun1}
\end{eqnarray} 
where $c(\{\nu\})$ are modified wavefunction amplitudes for each state $\{\nu\}$. After this modification adding  factors  $N(\{\nu\})$ the action of the Hamiltonian is expressed by integer numbers Eq. (\ref{eq:Schr22}). 

\begin{figure}[h!]
\centering
\includegraphics[width=8.5cm]{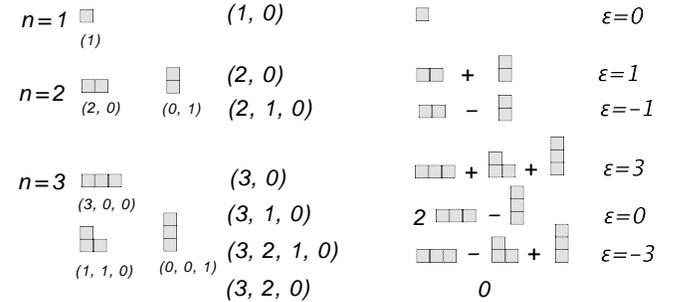}
\caption{Partitions, eigenstates and dimensionless anharmonic eigenenergies for principal quantum numbers  $n=1, 2$, and $3$ (See Ref. \cite{SI}, Sec. VI for detail).}
\label{fig:PartitIll}
\end{figure}

Since the harmonic energy is conserved in the resonant approximation ($[\widehat{H}_{res}, \widehat{H}_{0}]=0$,  see Eq. (\ref{eq:H})) the problem is reduced to the diagonalization of the dimensionless anharmonic interaction Hamiltonian $\widehat{V}$ and each contributing multiphonon state should possess the same harmonic energy $n/N$ satisfying the identity
\begin{eqnarray}
\sum_{i=1}^{n}\nu_{i}i = n. 
\label{eq:resEn}
\end{eqnarray}
Then each state can be characterized by its principal quantum number $n$.   Each selection of population numbers $\{\nu_{i}\}$ satisfying Eq. (\ref{eq:resEn}) corresponds to a certain integer partition of the number $n$ \cite{PartitionTheory} representing a way of writing n as a sum of positive integers.  For instance for the principal quantum number $n=3$ there exist three different  partitions ($3=1+1+1$, $3=2+1$, $3=3$, see Fig. \ref{fig:PartitIll}) corresponding to the phonon population numbers $(\nu_{1}, \nu_{2}, \nu_{3})= (3, 0, 0)$, $(1, 1, 0)$  and $(0, 0, 1)$.

Here we propose the algorithm to determine  eigenstates of the dimensionless Hamiltonian $\widehat{V}$ Eq. (\ref{eq:H}) and corresponding eigenenergies. One can describe the possible eigenstate using the sequence of $p+1$ integer numbers $\{k\}=(k_{0}, k_{1}, ...k_{p})$, such that  $k_{0}=n$ and $k_{p}=0$ ($n$ is the principal quantum number). The following rules determine the eigenstate and energy  corresponding to this sequence. 

{\bf 1}. Short sequences $(n, 0)$ correspond to the stationary states will all amplitudes equal to one ($c(\{\nu\})=1$, see Eq. (\ref{eq:EigFun1})) and energies $\epsilon_{(n, 0)}=n(n-1)/2$. This can be proved using direct substitution (cf. Eq. (\ref{eq:Schr22n0})). 

{\bf 2}. The eigenstate $c_{\{k\}}(\{\nu\})$ (if non-trivial) corresponding to the given sequence $\{k\}=(k_{0}, k_{1}, ...k_{p})$ can be defined using the eigenstate $c_{\{k_{-}\}}(\{m\})$ for the reduced sequence $\{k_{-}\}=(k_{1}, ...k_{p})$ with the principal quantum number $k_{1}$ ($\sum_{a}am_{a}=k_{1}$) obtained removing the zeroth term from the original sequence. The connection between two solutions and their energies can be expressed as  
\begin{eqnarray}
c_{\{k\}}(\{\nu\}) =\sum_{\{m\}}c_{\{k\}_{-}}(\{m\}) \psi_{\{m\}}(\{\nu\}),  
\nonumber\\
\epsilon_{\{k\}}=\epsilon_{\{k_{-}\}} + \frac{n(n-1)}{2}-nk_{1}+\frac{k_{1}(k_{1}-1)}{2}.
\label{eq:EigFun2}
\end{eqnarray}
where the summation is taken over all partitions $\{m\}$ of the number $k_{1}$ and the functions $\psi_{\{m\}}(\{\nu\})$ are given by the products of associated Laguerre polynomials \cite{Laguerre}
\begin{eqnarray}
\psi_{\{m\}}(\{\nu\})=\prod_{j=1}^{k_{2}}L_{m_{j}}^{(\nu_{j}-m_{j})}(j^{-1}).   
\label{eq:LaguerreProd}
\end{eqnarray}
To obtain the eigenstate described by the sequence $\{k\}$ this algorithm should be repeated $p$ times beginning with the sequence $(k_{p-1}, 0)$ corresponding to the all ones solution. 

{\bf 3}. If  the sequences $\{k\}$ are chosen strictly decreasing and satisfying the rule $k_{i-1}-k_{i} \geq k_{i}-k_{i+1}$ then the number of sequences is equal to the number of partitions. The numerical studies show that up to the maximum accessed principal quantum number $n=25$ the eigenstates generated using this sequences following the above algorithm form the complete basis of eigenstates all normalized by one and orthogonal to each other. Unfortunately, we cannot give a general proof of this statement for arbitrarily %principal quantum number  
$n$ though the normalization by $1$ is proved for some groups of generated states \cite{SI} Secs. II, V. The use of basis functions with smaller quantum numbers $k_{1}<n$ to describe the partitions of larger number $n$ does not conflict with the completeness of the basis because the populations numbers are dependent of each other, i. e. they are bound by the ``harmonic energy conservation law" Eq. (\ref{eq:resEn}). 

The dimensionless energy of the eigenstate obtained repeating $p-1$ times the iteration procedure Eq. (\ref{eq:EigFun2}) for a certain sequence $\{k\}$  is given by 
\begin{eqnarray}
\epsilon(\{k\})=-\frac{n(n-1)}{2}+\sum_{i=0}^{p-1}\left[k_{i}(k_{i}-1)-k_{i}k_{i+1}\right].
\label{eq:EigEn}
\end{eqnarray} 
%where the energies $\epsilon(\{k\})$ represent eigenenergies of the dimensionless anharmonic interaction $\widehat{V}$. 

\begin{figure}[h!]
\centering
\includegraphics[width=9cm]{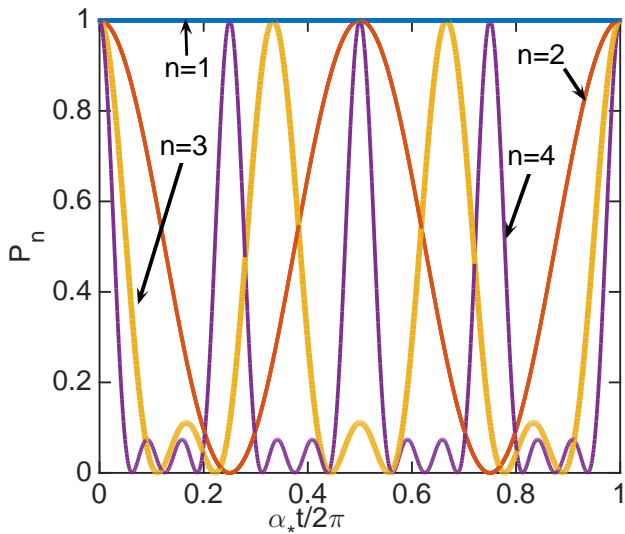}
\caption{Time evolution of the probability for the system to remain in its initial single phonon state with the principal quantum number $n$ (see Eq. (\ref{eq:Pj})).}% we set $\alpha t/N\hbar =2\pi$). }
\label{fig:PhL}
\end{figure}

Using these eigenstates one can describe the exact time evolution of the single phonon state assuming that at time  $t=0$ there was only one phonon in the $n^{th}$ harmonic state with the energy $n/N$ (cf. Eq. (\ref{eq:H})). %This state can be expanded over the $n$ eigenstates $\psi(p)$ determined by sequences of strictly decreasing integer numbers $n, p-1, p-2... 0$, enumerated by the integer number $p$ changing from $1$ to $n$, having energies $\epsilon(p)=-n(n-1)/2+n(n-p)$  Eq. (\ref{eq:EigEn}) and overlap integral $1/\sqrt{n}$ with this single phonon state  (see derivation below and in the Supplementary Materials \cite{SI}). 
It can be shown (see below Eq. (\ref{eq:psim})) that the probability that the  system remains in the single phonon state oscillates with the time as (see Fig. \ref{fig:PhL})
\begin{eqnarray}
P_{n}(t)=\frac{\sin^2\left(\alpha_{*} n^2 t\right)}{n^2\sin^2\left(\alpha_{*}nt\right)}, ~\alpha_{*}=\frac{\sqrt{2}\pi\alpha}{4N^2\hbar}.  
\label{eq:Pj}
\end{eqnarray}
These oscillations and the oscillation period dependence on the anharmonic interaction  and the system size are similar to the behaviors discovered  in Ref. \cite{FPUclassic}. 

To illustrate the proposed algorithm we show in  Fig. \ref{fig:PartitIll} all  eigenstates for $n=1$, $2$ and $3$ obtained using strictly decreasing sequences of quantum numbers $k$ (for detail see Ref. \cite{SI}, Sec. VI). Basis states divided by corresponding normalization factors (see Eq. (\ref{eq:EigFun1})) are represented by diagrams \cite{PartitionTheory}. For $n=1, 2$ the strictly decreasing sequences represent the complete basis of eigenstates of the problem. For $n=3$ three sequences $(3, 0)$, $(3, 1, 0)$ and $(3, 2, 1, 0)$ represent the complete basis set in agreement with the proposed algorithm.  The sequence $(3, 2, 0)$ leads to the zero wavefunction and there are many more of such sequences  for larger $n$ among strictly decreasing sequences. This is because the total number $2^{n}$ of strictly decreasing sequences $\{k\}$ is much greater than the total number of partitions depending on  $n$ as \cite{PartitionTheory}
\begin{eqnarray}
p(n) \approx \frac{1}{4\sqrt{3}n}\exp\left(\pi\sqrt{\frac{2n}{3}}\right).% \ll 2^{n}. 
\label{eq:partNum}
\end{eqnarray}

Consider the derivation of the results, described above. To find eigenstates of  Eq. (\ref{eq:H}) we can use the wavefunction in the form Eq. (\ref{eq:EigFun1}) for some principal quantum number $n$ expressing the harmonic part of energy.  The Schr\"odinger equation  for the dimensionless anharmonic Hamiltonian $\widehat{V}$ (cf. Eq. (\ref{eq:H})) can be written as 
\begin{eqnarray}
\epsilon c_{\{\nu\}}=\frac{1}{2}\sum_{a, b}ab\nu_{a}(\nu_{b}-\delta_{ab})\widehat{x}_{a}^{-}\widehat{x}_{b}^{-}\widehat{x}_{a+b}^{+}c_{\{\nu\}}
\nonumber\\
+\frac{1}{2}\sum_{a, b}(a+b)\nu_{a+b}\widehat{x}_{a}^{+}\widehat{x}_{b}^{+}\widehat{x}_{a+b}^{-}c_{\{\nu\}}.
\label{eq:Schr22}
\end{eqnarray}  
The introduced $\widehat{x}$-operators $\widehat{x}_{a}^{\pm}$ raise or lower the population index $\nu_{a}$ of the amplitude  $c_{\{\nu\}}$ by $1$ (for example for $n=3$ one can express the wavefunction amplitudes as $c_{\{\nu\}}=c_{\nu_{1},\nu_{2},\nu_{3}}$ and then  $x_{2}^{\pm}c_{\nu_{1},\nu_{2},\nu_{3}}=c_{\nu_{1},\nu_{2}\pm 1,\nu_{3}}$). Population numbers cannot be negative; fortunately the related terms disappear in Eq. (\ref{eq:Schr22}) because of the zero factor $\nu_{a}\nu_{b}$ so there is no need to care about them. 

We begin with the consideration of the solution determined by the sequence $(n, 0)$ which is claimed to be $c_{\{\nu\}}=1$ for all partitions $\{\nu\}$ and   it should have the eigenenergy $\epsilon=n(n-1)/2$ Eq. (\ref{eq:EigEn}). Indeed, assuming all identical amplitudes $c_{\{\nu\}}$ one can rewrite Eq. (\ref{eq:Schr22}) for some specific partition $\{\nu\}$ as      
\begin{eqnarray}
\epsilon =\frac{1}{2}\left[ \sum_{a, b}ab\nu_{a}\nu_{b} -\sum_{a}a\nu_{a}\right].
\label{eq:Schr22n0}
\end{eqnarray} 
Since for each partition $\{\nu\}$ in Eq. (\ref{eq:EigFun1})  one has $\sum_{a}a\nu_{a}=n$ (see Eq. (\ref{eq:resEn})) we got $\epsilon=n(n-1)/2$ as in Eq. (\ref{eq:EigEn}). %Moreover this state is normalized by $1$ \cite{SI}. 

The recursive  algorithm of finding solutions described above is based on the following property of solutions of Eq. (\ref{eq:Schr22}). If the set of amplitudes  $d_{\{m\}}$ for the principal quantum number $m$ describes the solution with the energy $\epsilon_{d}$ it can be used to generate another solution $c_{\{\nu\}}$ (if it is non-trivial) with the principal quantum number $n$ and eigenenergy $\epsilon_{c}$ defined using Eq. (\ref{eq:LaguerreProd}) as 
\begin{eqnarray}
c_{\{\nu\}} =\sum_{\{m\}}d_{\{m\}}\psi_{\{m\}}(\{\nu\}),   
\nonumber\\
\epsilon_{c}=\epsilon_{d} + \frac{n(n-1)}{2}-nm+\frac{m(m-1)}{2}.
\label{eq:EigFun3}
\end{eqnarray}
This result can be derived substituting the solution in the form of Eq. (\ref{eq:EigFun3}) into the Schr\"odinger equation Eq. (\ref{eq:Schr22}) and simplifying it using  the properties of Laguerre polynomials \cite{Laguerre} (see  Sec. III in Ref.  \cite{SI} for detail).  The algorithm Eq. (\ref{eq:EigFun3}) of finding eigenstates and eigenenergies is implemented in Matlab codes which are the part of Ref. \cite{SI} (Sec. VII in the text and the code files) and the reader is strongly encouraged to use them and Sec. VI in Ref. \cite{SI} to verify the relevance of the proposed algorithm.

Choosing $k_{0}=n$, $k_{1}=m$ one  can identify the expression of amplitudes $c_{\{\nu\}}$ in terms of amplitudes $d_{\{m\}}$ as the first iteration step Eq. (\ref{eq:EigFun2}) which can be repeated (backwards) arbitrarily number of times until the termination at $k_{p}=0$ where all amplitudes should be set equal unity as described above. The corresponding evolution of energy  by the set of shifts Eq. (\ref{eq:EigFun3})  leads to Eq. (\ref{eq:EigEn}). Thus each integer number sequence defines the eigenstate and eigenenergy of the problem if this eigenstate is nontrivial.

The amplitudes $c_{\{k\}}({\{\nu\}})$ for the specific system eigenstate determined by the  sequence $\{k\}$ of $p+1$ quantum numbers %(remember that $1\leq p \leq n$, $k_{0}=n$, $k_{p}=0$) 
and taken for the specific population number set $\{\nu\}$ can be expressed by means of the generating function calculated using Eq. (\ref{eq:EigFun2}) in the form (see Ref. \cite{SI}, Sec. IV) 
\begin{widetext}  
\begin{eqnarray}
G(\{\nu\}, \{y\})=\sum_{\{k\}}c_{\{k\}}({\{\nu\}})k_{1}^{y_{1}}k_{2}^{y_{2}}...k_{p-1}^{y_{p-1}}
=\prod_{f, g (1<f\leq g<p)}\left[1-\prod_{i=f}^{g}y_{i}\right]\prod_{j=1}^{n}\left[1+y_{1}^{j}+%(y_{1}y_{2})^k+
...(y_{1}..y_{p-1})^j\right]^{\nu_{j}}. 
\label{eq:wfderfunAns}
\end{eqnarray}
\end{widetext}
Wavefunction amplitudes for the sequence $\{k\}$ are expressed by the coefficients of the generating function expansion with the term containing the product $y_{1}^{k_{1}}y_{2}^{k_{2}}...y_{p-1}^{k_{p-1}}$. 

Eq. (\ref{eq:wfderfunAns}) can be used to define all states having non-zero overlap with the single phonon state $|n>$ represented by the population number sequence $\nu_{k}=\delta_{kn}$, which is needed to describe the time evolution of this state Eq. (\ref{eq:Pj}). We consider only strictly decreasing sequences $\{k\}$ which is sufficient to get a complete expansion of the initial state over eigenstates of the problem as shown below. 
Then for the state $\nu_{k}=\delta_{kn}$  one can leave only unity from the second factor in the right hand side of Eq. (\ref{eq:wfderfunAns}) because  otherwise the power of some of variables $y_{i}$ will approach or exceed $n$ which is not acceptable since these powers should form strictly decreasing sequence $n>k_{1}>k_{2}>k_{3}...>k_{p}=0$. The only acceptable choice of the contribution from  the first product satisfying the sequence selection  requirements can be made taking the composite product $y_{1}\times (y_{1}y_{2})\times (y_{1}y_{2}y_{3})...\times (y_{1}y_{2}...y_{p-1})$. The related sequence is given by $(n, p-1, p-2, p-3,...0)$ and it determines the stationary state $\Psi_{p}$ characterized by the energy $\epsilon_{p}=(-n(n-1)/2+n(n-p))$ (cf. Eq. (\ref{eq:EigEn})).  There are $n$ such sequences and associated eigenstates $\Psi_{p}$ determined by the integer number $p$ changing from $1$ to $n$ (cf. the solutions for eigenstates in Ref. \cite{SI}, Sec. VI for $n\leq 5$). In all cases (see Ref. \cite{SI}, Sec.  V) the absolute value of the amplitude of the wavefunction in the single phonon state is equal unity, and it can be recalculated as $1/\sqrt{n}$ for the normalized by $1$ representation Eq. (\ref{eq:EigFun1}). Thus we found $n$ contributing eigenstates having the overlap integral $c_{p}=1/\sqrt{n}$ with the single phonon state of interest. The expansion of the single phonon state over the basis of those $n$ states is complete since $nc_{p}^2=1$.

Assume that at time $t=0$ the system is in a single phonon state $|n>$. Then the time evolution of the probability $P_{n}(t)$ to find the system in this specific state can be evaluated expanding this state over the previously established set of $n$ eigenstates $\Psi_{p}$ as 
\begin{eqnarray}
P_{n}(t)=\left|\sum_{p=1}^{n} e^{-i\frac{\alpha_{*}\epsilon_{p}t}{\hbar}}|<\Psi_{p}|n>|^2\right|^2, ~ \alpha_{*}=\frac{\sqrt{2}\alpha}{4N^2\hbar}. 
\label{eq:psim}
\end{eqnarray}
Since all overlap matrix elements are identical so that  $|<\Psi_{p}|n>|^2=1/n$ the sum in Eq. (\ref{eq:psim})  takes the form of the geometric series $\sum_{p=1}^{n}e^{2i\alpha_{*} npt}$. The evaluation of this geometric series results in Eq. (\ref{eq:Pj}).  

Consider some properties of eigenstates of the problem. Since Eq. (\ref{eq:Schr22}) has a symmetry with respect to the transformation $\epsilon \rightarrow -\epsilon$, $c_{\{\nu\}}\rightarrow c_{\{\nu\}}\cdot (-1)^{\sum_{i}\nu_{i}}$ all eigenstates with non-zero energy enter in pairs (this conclusion is illustrated in Ref. \cite{SI} for $n\leq 5$). Particularly the all ones state determined by the sequence $(n, 0)$ possessing the energy $n(n-1)/2$ has a corresponding state determined by the sequence $(n, n-1, n-2...0)$ possessing the opposite energy (see Eq. (\ref{eq:EigEn}), Ref. \cite{SI}). The first states possesses the maximum of energy because it has all positive amplitudes, while the anharmonic Hamiltonian has all positive matrix elements. Consequently the second state possesses the energy minimum and energies of all other states belong to the domain $(-n(n-1)/2, n(n-1)/2)$. 
Since all energies are expected to be expressed by integer numbers Eq. (\ref{eq:EigEn}) and the number of states (partitions, Eq. (\ref{eq:partNum})) grows with the principal quantum number $n$ faster than any power of $n$  the strong degeneracy is expected at large $n$, reflecting the integrability of the problem.

The fourth order anharmonic interaction can be introduced within the resonant approach similarly to Eq. (\ref{eq:H}). It will represent the $\beta$ FPU model characterized by some constant $\beta$ of the fourth order anharmonic interaction.  The preliminary numerical study of this problem does not lead to the analytical solution; yet, the small modification of the resonant $\beta$ model by adding to the original Hamiltonian the diagonal term in phonon population numbers proportional to the expression
\begin{eqnarray}
\sum_{a}\left[a^2\nu_{a}(1+\nu_{a})-\frac{1}{3}a^3\nu_{a}\right]
\frac{1}{12}
\sum_{a}a^3\nu_{a}
\label{eq:FourthOrder}
\end{eqnarray}
makes the problem eigenstates identical to those of the resonant FPU $\alpha$ model. The accurate analysis of the $\beta$ FPU problem will be performed separately. 

In addition to vibrational energy transport the obtained solution can be relevant for quantum informatics because of its connection to the number  theory \cite{PartitionTheory}. Therefore its realizations employing interacting Josephson junctions or cold atoms \cite{Flach,Lukin2}  is of interest and the model Eq. (\ref{eq:H}) can be hopefully implemented there with a high accuracy. 

In realistic polymers the breakdown of integrable behavior and transition to chaos are possible due to the omitted ``non-trivial resonances'' (due to high order processes messing up positive and negative wavevectors) inevitably leading to the ergodic behavior in a classical system according to Ref. \cite{PNASRecent}. We hope that  in a quantum system an integrable state can be stable because of the phase space discreteness  \cite{LeitnerArnoldDiff}.  This problem needs a separate  study. 

Even at small phonon energy $n\sim 1$ the deviation of the phonon dispersion law from the linear one $1/N^3$ Ref. \cite{SI}, Sec. I can destroy Fermi resonances when it exceeds the Fermi resonance amplitude $\alpha/N^2$ Eq. (\ref{eq:H}).  Consequently, the proposed theory can be applicable for a sufficiently large system size $N>1/\alpha$. For organic polymers $\alpha\sim 0.1$, and the regime $N>1/\alpha \sim 10$ is quite accessible. The long wavelength limit requires a typical phonon energy $n^{1/2}/N$ ($n$ is a principal quantum number and the typical phonon energy is taken as a thermal energy corresponding to the total energy $n/N$ \cite{TextPh}) to be less than the Debye energy, which is of order of $1$ within the FPU model. More accurate analysis of applicability limits for the present solution requires separate investigation. The obtained solution should not be very sensitive to defects %provided that the energy expressed by the principal quantum number $n$ is sufficiently small since the scattering by disordering disappears 
in the long-wavelength low energy limit \cite{LeibovitzLocalNonlin,DharRev,David1,abPhR}. %Moreover the anharmonic interaction will reduce the emergence of Anderson localization because of the increase of the effective number of transport channels which can be potentially described using the present results.

Thus the family of analytical solutions is found for eigenstates  of the quantum mechanical many-body problem of one dimensional acoustic vibrations coupled by the resonant anharmonic interactions. According to the numerical studies this family forms the complete set of eigenstates. Eigenstates are described by sequences of quantum numbers possibly representing the integrals of motion of unclear nature which calls for  further theoretical studies. Practically the present model on a one hand  is closely related to the vibrational energy transport in molecular chains and on the other hand it is connected to the number theory thus having a potential interest in quantum informatics. Therefore we hope that this work will stimulate experimental efforts to implement the present model using cold atoms and/or Josephson junctions and contribute to understanding the thermal conductivity of polymers.  

This work is supported by the National Science Foundation (CHE-1462075).  

% u^4 

% Scattering, thermal conductivity

% Conclusions and consequences
% We found quantum mechanical solution, integrals of motions expressed by quantum numbers still need to be understood 
% Interesting universality class, awaits complete understanding and connection to solitons

\newpage
\begin{widetext}
\begin{center}
{\Large{\bf Supplementary Materials}}
\end{center}

\section{Derivation of the Resonant anharmonic Hamiltonian}

We begin with the derivation of the resonant Hamiltonian for the FPU $\alpha$ problem describing anharmonic vibrations of atoms forming periodic chain. Remember that $\alpha$ problem  involves the third order anharmonic interactions only.  The system Hamiltonian is defined following the seminal work of Fermi, Pasta and Ulam \cite{FPUclassic} as  
\begin{eqnarray}
\widehat{H} = \frac{1}{2}\sum_{i=1}^{N}p_{i}^{2}+\frac{1}{2}\sum_{i=1}^{N-1}(x_{i}-x_{i+1})^2 + \frac{1}{2}(x_{N}-x_{1})^2+
\nonumber\\
+\frac{\alpha}{3}\sum_{i=1}^{N-1}(x_{i}-x_{i+1})^3 + \frac{\alpha}{3}(x_{N}-x_{1})^3.
\label{eq:HsimlFourth}
\end{eqnarray}
This  Hamiltonian includes harmonic (first term) and third order anharmonic (second term) interactions.

Normal modes diagonalizing the harmonic part of the Hamiltonian Eq. (\ref{eq:HsimlFourth}) can be introduced  as periodic waves or phonons (assuming even $N$ without the lack of generality) 
\begin{eqnarray}
x_{n} = \frac{1}{\sqrt{N}}\sum_{q}e^{iqn}u_{q}, ~ 
%\nonumber\\
u_{q} = \frac{1}{\sqrt{N}}\sum_{n=1}^{N}e^{-iqn}x_{n}, 
\nonumber\\
q=\frac{2\pi m}{N}, ~ m= -\frac{N}{2}+1, ... \frac{N}{2}.   
\label{eq:NormModes}
\end{eqnarray}
Similar transformation should be performed for atomic momenta. 

Using these new coordinates and momenta one can reexpress the Hamiltonian Eq. (\ref{eq:HsimlFourth}) as 
\begin{eqnarray}
\widehat{H} = \frac{1}{2}\sum_{q\geq 0}\left(p_{q}p_{-q}+\omega_{q}^2u_{q}u_{-q}\right)+
\nonumber\\
+\frac{\alpha}{3\sqrt{N}}\sum_{q_{1}, q_{2}, q_{3}}\Delta_{q_{1}+q_{2}+q_{3}}(1-e^{iq_{1}})(1-e^{iq_{2}})(1-e^{iq_{3}})u_{q_{1}}u_{q_{2}}u_{q_{3}}, 
\nonumber\\
\omega(q)=2\mid\sin(q/2)\mid,
\nonumber\\
\Delta_{q_{1}+q_{2}+q_{3}}=\sum_{n=-\infty}^{\infty}\delta_{m_{1}+m_{2}+m_{3},nN},     
\label{eq:HsimlFourthNorm}
\end{eqnarray}
where $\omega(q)$ represents the vibrational frequency of a phonon with the wavevector $q$, the $\delta$ symbol, $\delta_{m,n}=1$ for $m=n$ and $0$ otherwise, stands for the Kronecker symbol and $m_{i}$ represents the integer quantization number corresponding to the wavevector $q_{i}$ Eq. (\ref{eq:NormModes}). Since at $q \ll 1$ one has $\omega \approx q$ the speed of sound is equal to unity within the FPU $\alpha$ model. 

It is convenient to express the Hamiltonian in terms of creation and annihilation operators of vibrational modes defined as
\begin{eqnarray}
u_{q}=\sqrt{\frac{\hbar}{2\omega(q)}}(b^{+}_{q}+b_{-q}), ~ p_{q}=i\sqrt{\frac{\hbar\omega(q)}{2}}(b^{+}_{q}-b_{-q}).
\label{eq:bb+}
\end{eqnarray}
Consequently one can represent the harmonic Hamiltonian in its standard diagonal form 
\begin{eqnarray}
\widehat{H}_{0}=\sum_{q}\hbar\omega_{q}(b_{q}^{+}b_{q}+1/2),
\label{eq:Hharm}
\end{eqnarray}
while anharmonic interactions can be expressed as %(only resonant interactions capable to conserve energy are included)
\begin{eqnarray}
\widehat{V}_{3} = i\frac{\alpha\hbar^{\frac{3}{2}}}{\sqrt{N}}\sum_{q_{1}, q_{2}, q_{3}}\Delta_{q_{1}+q_{2}+q_{3}}e^{i\frac{q_{1}+q_{2}+q_{3}}{2}}\frac{\sin(q_{1}/2)\sin(q_{2}/2)\sin(q_{3}/2)}{\sqrt{|\sin(q_{1}/2)\sin(q_{2}/2)\sin(q_{3}/2)|}}\left(b_{q_{1}}^{+}b_{q_{2}}^{+}b_{-q_{3}} -  b_{-q_{3}}^{+}b_{q_{1}}b_{q_{2}}\right)+
\nonumber\\
+i\frac{\alpha\hbar^{\frac{3}{2}}}{3\sqrt{N}}\sum_{q_{1}, q_{2}, q_{3}}\Delta_{q_{1}+q_{2}+q_{3}}e^{i\frac{q_{1}+q_{2}+q_{3}}{2}}\frac{\sin(q_{1}/2)\sin(q_{2}/2)\sin(q_{3}/2)}{\sqrt{|\sin(q_{1}/2)\sin(q_{2}/2)\sin(q_{3}/2)|}}\left(b_{q_{1}}^{+}b_{q_{2}}^{+}b_{q_{3}}^{+} -  b_{q_{3}}b_{q_{1}}b_{q_{2}}\right).
\label{eq:HFourthModbb1}
\end{eqnarray}

Assuming that the only low energy vibrations are considered $q_{1}, q_{2}. q_{3} \ll 1$ (i. e. $n_{1}, n_{2}. n_{3} \ll N$ ) we can restrict the summation in the $\Delta$ symbol definition in Eq. (\ref{eq:HsimlFourthNorm}) to the case $n=0$ assuming the conservation of quasi-momentum and set $sin(q/2)\approx q/2$ everywhere in Eqs. (\ref{eq:Hharm}) and  (\ref{eq:HFourthModbb1}). Then the Hamiltonian takes the approximate form 
\begin{eqnarray}
\widehat{H} = \frac{h}{N}\sum_{n=-\infty}^{\infty}\mid n\mid(b_{n}^{+}b_{n}+1/2)+
\nonumber\\
-i\frac{\alpha h^{\frac{3}{2}}}{\sqrt{8}N^2}\sum_{m, n}\frac{mn(m+n)}{\sqrt{|mn(m+n)|}}\left(b_{m}^{+}b_{n}^{+}b_{m+n} -  b_{m+n}^{+}b_{m}b_{n}\right)-
\nonumber\\
-i\frac{\alpha h^{\frac{3}{2}}}{3\sqrt{8}N^2}\sum_{m, n}\frac{mn(m+n)}{\sqrt{|mn(m+n)|}}\left(b_{-m-n}^{+}b_{m}^{+}b_{n}^{+} -  b_{-m-n}b_{m}b_{n}\right).
\label{eq:HFourthModbb+}
\end{eqnarray}

Finally we leave only ``fully" resonant processes conserving both energy and quasimomenta and perform the unitary transformation $b_{n}^{\dagger} \rightarrow ib_{n}^{\dagger}$, $b_{n} \rightarrow -ib_{n}$. Then the Hamiltonian can be separated into positive and negative wavevector $n, m$ parts which can be treated separately. The positive wavevector part can be written as 
\begin{eqnarray}
\widehat{H}_{res} = \widehat{H}_{0}+\frac{\alpha h^{\frac{3}{2}}}{2\sqrt{2}N^2}\widehat{V}; ~ \widehat{H}_{0}= \frac{h}{N}\sum_{n>0}n(b_{n}^{+}b_{n}+1/2); 
\nonumber\\
%\widehat{V}=\frac{\alpha h^{\frac{3}{2}}}{2\sqrt{2}N^2}\sum_{m, n >0}\sqrt{mn(m+n)}\left(b_{m}^{+}b_{n}^{+}b_{m+n} +  b_{m+n}^{+}b_{m}b_{n}\right).
\widehat{V}=\frac{1}{2}\sum_{m, n >0}\sqrt{mn(m+n)}\left(b_{m}^{+}b_{n}^{+}b_{m+n} +  b_{m+n}^{+}b_{m}b_{n}\right).
\label{eq:Hfin}
\end{eqnarray}
This Hamiltonian is studied within the main body of the manuscript. Since the harmonic energy is conserved (the harmonic Hamiltonian $\widehat{H}_{0}$ commutes with the whole Hamiltonian) the problem can be reduced to the diagonalization of the anharmonic dimensionless Hamiltonian $\widehat{V}$. 

The correction to the linear dispersion law for the phonon frequency Eq. (\ref{eq:HsimlFourthNorm}) for the given quantum number $n$ can be estimated as $\delta\omega \sim (n/N)^3$. This correction is always smaller than the third order anharmonic interaction Eq. (\ref{eq:HFourthModbb+}) at sufficiently  large system size $N$ because the latter interaction decreases with this size as $N^{-2}$. 

\section{Normalization of the solution for the sequences ($n$, 0).}
\label{sec:Normn0}

For each partition $\{\nu\}$ of the number $n$ representing the corresponding sequence of population numbers the squared amplitude $a_{\{\nu\}}$ of the wavefunction for the state determined by the sequence $(n, 0)$ is given by (we assumed all amplitudes $c_{\{\nu\}}=1$)
\begin{eqnarray}
|a_{\{\nu\}}|^2=\frac{|c_{\{\nu\}}|^2}{\prod_{i=1}^{n}i^{\nu_{i}}\nu_{i}!}=\frac{1}{\prod_{i=1}^{n}i^{\nu_{i}}\nu_{i}!}. 
\label{eq:n0norm}
\end{eqnarray}
One can calculate the sum of squared amplitudes $a_{\{\nu\}}$ over all possible partitions using the generating function $G(x)$ defined as 
\begin{eqnarray}
G(x)=\sum_{\{\nu\}}|a_{\{\nu\}}|^2 x^{\sum_{k}k\nu_{k}}, 
\label{eq:Genn0}
\end{eqnarray}
where the summation is over all possible partitions $\{\nu\}$ for all integer numbers $n$. Then this sum can be represented by the power series $G(x)=\sum_{k}N_{k}x^{k}$, where the coefficients $N_{k}$ are defined as the normalization factors for each principal quantum number $k$, i. e.  
\begin{eqnarray}
N_{k}=\sum_{\{\nu\}_{k}}|a_{\{\nu\}_{k}}|^2. 
\label{eq:Coefn0}
\end{eqnarray}

Using the definition of the amplitudes Eq. (\ref{eq:n0norm}) one can evaluate the generating function as 
\begin{eqnarray}
G(x)=\sum_{\{\nu\}}|a_{\{\nu\}}|^2 x^{\sum_{k}k\nu_{k}}=\exp\left(\sum_{k}\frac{x^{k}}{k}\right)=\frac{1}{1-x}.  
\label{eq:Genn0a}
\end{eqnarray}
Since all coefficients in the expansion of Eq. (\ref{eq:Genn0a}) with respect to the powers of $x$ are equal to $1$ we can conclude that all eigenstates $(n, 0)$ are normalized by $1$ for arbitrarily $n$. 

Using a similar method one can also prove the normalization by one for the quantum number  sequence ($n, m, m-1,...0$) with $m<n$  (see Sec. \ref{sec:Normp}). Numerical probes show that the normalization by one holds for all non-trivial eigenstates generated from  sequences determined by the algorithm formulated within the main text. Yet we cannot prove this statement in a general case.

\section{Derivation of the recursive equation}

The original Schr\"odinger equation for the modified wavefunction amplitudes $c_{\{\nu\}}=a_{\{\nu\}}\sqrt{N_{\{\nu\}}}$ (Eq. (\ref{eq:Coefn0})) with respect to the anharmonic Hamiltonian $\widehat{V}$  has the form 
\begin{eqnarray}
\epsilon c_{\{\nu\}}=\widehat{h}c_{\{\nu\}}=
%\nonumber\\
\frac{1}{2}\sum_{a, b}ab\nu_{a}(\nu_{b}-\delta_{ab})\widehat{x}_{a}^{-}\widehat{x}_{b}^{-}\widehat{x}_{a+b}^{+}c_{\{\nu\}}+
\nonumber\\
+\frac{1}{2}\sum_{a, b}(a+b)\nu_{a+b}\widehat{x}_{a}^{+}\widehat{x}_{b}^{+}\widehat{x}_{a+b}^{-}c_{\{\nu\}},
\label{eq:OrigPr}
\end{eqnarray} 
where the operator $\widehat{h}$ describes the action of the anharmonic Hamiltonian $\widehat{V}$  in the representation of modified amplitudes  $c_{\{\nu\}}$. 
Remember that %the wavefunction amplitudes $a(\{\nu\})$ normalized by one are related to the amplitudes  $c(\{\nu\})$ by means of Eq.(\ref{eq:n0norm}) and the 
$\widehat{x}$-operators $\widehat{x}_{a}^{\pm}$ raise or lower the population index $\nu_{a}$ of the amplitude $c(\{\nu\})$ by $1$. 

Assume that the amplitudes $c_{\{\nu\}}$ are expanded in terms of the basis set composed by Laguerre polynomial products 
\begin{eqnarray}
\psi_{\{m\}}(\{\nu\})=\prod_{j=1}^{k_{2}}L_{m_{j}}^{(\nu_{j}-m_{j})}(j^{-1}),
\label{eq:LaguerreProdApp}
\end{eqnarray} 
 as 
\begin{eqnarray}
c_{\{\nu\}}=\sum_{\{m\}}d_{\{m\}}\psi_{\{m\}}(\{\nu\})\psi_{I}, ~ \psi_{I}=\delta_{\sum_{i}i\nu_{i},n},
\label{eq:SolAnz}
\end{eqnarray}
where  the Kronecker symbol $\delta_{\sum_{i}i\nu_{i},n}$ is equal to one for the population number set $\{\nu_{i}\}$ satisfying the  conservation law 
\begin{eqnarray}
\sum_{i=1}^{n}i\nu_{i}=n. 
\label{eq:Ident1}
\end{eqnarray}
%Eq. (\ref{eq:Ident1}). 
This symbol defines the all unities wave function corresponding to the sequence $(n, 0)$.  
%\begin{eqnarray}
%\sum_{i}i\nu_{i}=n,  
%\label{eq:Ident1}
%\end{eqnarray}

We are going to show that the amplitudes $d_{\{m\}}$ in Eq. (\ref{eq:SolAnz}) can be chosen in the way that they differ from zero only for sequences $\{m\}$  representing integer partitions of some number $m$ suggesting 
\begin{eqnarray}
\sum_{i}im_{i}=m.  
\label{eq:Ident2}
\end{eqnarray}
Moreover these amplitudes satisfy the equation 
\begin{eqnarray}
\left(\epsilon - \frac{n(n-1)}{2}+nm-\frac{m(m-1)}{2}\right)d_{\{m\}}=
\nonumber\\
=\frac{1}{2}\sum_{a, b}ab\nu_{a}(\nu_{b}-\delta_{ab})\widehat{y}_{a}^{-}\widehat{y}_{b}^{-}\widehat{y}_{a+b}^{+}d_{\{m\}}
%\nonumber\\
+\frac{1}{2}\sum_{a, b}(a+b)\nu_{a+b}\widehat{y}_{a}^{+}\widehat{y}_{b}^{+}\widehat{y}_{a+b}^{-}d_{\{m\}}
\label{eq:RecEq}
\end{eqnarray} 
where raising and lowering operators $\widehat{y}^{\pm}$ act on the indices $m$.  
 This equation is almost identical to Eq. (\ref{eq:OrigPr}) except for the redefinition of energy that determines the recursive procedure defining the energy of each specific eigenstate in terms of the set of the associated quantum numbers. 
 %One can choose $m<n$ because the solutions with $m\geq n$ are linear dependent of solutions with $m<n$ (see Sec. \ref{sec:maxeffpow}).

To derive this equation one can seek the solution in the form 
\begin{eqnarray}
\Psi=\sum d_{\{m\}}\psi_{\{m\}}(\{\nu\}), 
\label{eq:Anz2}
\end{eqnarray}
where the sum is taken over all integer partitions of a certain number $m$ Eq. (\ref{eq:Ident2}). %Remember %that the basis functions  $\psi_{\{m\}}(\{\nu\})$ are defined in terms of the products of Laguerre polynomials as 
%\begin{eqnarray}
%\psi_{\{m\}}(\{\nu\})=\prod_{j=1}^{m}L_{m_{j}}^{(\nu_{j}-m_{j})}(j^{-1}).    
%\label{eq:LaguerreProdApp}
%\end{eqnarray}

The wavefunction amplitudes Eq. (\ref{eq:SolAnz}) can be expressed as the results of the action of operators $\widehat{\psi}=\psi_{\{m\}}(\{\widehat{n}\})$ on the unit wavefunction $\psi_{I}$ Eq. (\ref{eq:SolAnz}). One can then represent the action of the Hamiltonian in Eq. (\ref{eq:OrigPr}) as 
\begin{eqnarray}
\widehat{h} \widehat{\psi}\psi_{I}=\widehat{\psi} \widehat{h}\psi_{I}+[\widehat{h}, \widehat{\psi}]\psi_{I}.
\label{eq:CommNeed}
\end{eqnarray}
It was shown in the main body of the manuscript that the state $\psi_{I}$ with all amplitudes $c$ equal $1$ is the eigenstate of the problem with the eigenenergy $n(n-1)/2$ so that $\widehat{h}\psi_{I}=n(n-1)/2 \psi_{I}$. Consider the commutator $[\widehat{h}, \widehat{\psi}]$ related part of the problem.

Since we are interested in the Laguerre polynomial dependence  of population numbers $n$ it is convenient to introduce the different notations for them as
\begin{eqnarray}
P_{m_{c}}^{c}(\nu_{c})=L_{m_{c}}^{(\nu_{c}-m_{c})}(1/c),
\label{eq:LagRepl}
\end{eqnarray}
which is easier to follow during the derivation below.

The commutation rules that can be used to evaluate the expressions in Eq. (\ref{eq:CommNeed}) can be summarized as following 
\begin{eqnarray}
[x_{a}^{-}, P_{m_{c}}^{c}(\nu_{c})]=\delta_{ac} (P_{m_{c}}^{c}(\nu_{c}-1)-P_{m_{c}}^{c}(\nu_{c})) x_{a}^{-}=-\delta_{ac} P_{m_{c}-1}^{c}(\nu_{c}-1)x_{a}^{-};
\nonumber\\
\lbrack x_{b}^{-} , \lbrack x_{a}^{-}, P_{m_{c}}^{c}(\nu_{c})\rbrack\rbrack
=\delta_{ac}\delta_{bc}P_{m_{c}-2}^{c}(\nu_{c}-2); 
\nonumber\\
\lbrack x_{a}^{+}, P_{m_{c}}^{c}(\nu_{c})\rbrack=\delta_{ac} (P_{m_{c}}^{c}(\nu_{c}+1)-P_{m_{c}}^{c}(\nu_{c})) x_{a}^{+}=\delta_{ac} P_{m_{c}-1}^{c}(\nu_{c})x_{a}^{+};
\nonumber\\
\lbrack x_{b}^{+}, \lbrack x_{a}^{+}, P_{m_{c}}^{c}(\nu_{c})\rbrack\rbrack=\delta_{ac}\delta_{bc}P_{m_{c}-2}^{c}(\nu_{c}).  
\label{eq:CommRules1}
\end{eqnarray} 
These rules are based on identities for Laguerre polynomials \cite{Laguerre} that have been used during the derivation of commutators in Eq. (\ref{eq:CommRules1}). These identities can be rewritten in the notations of Eq. (\ref{eq:LagRepl}) in the form 
\begin{eqnarray}
P_{m_{c}}^{c}(\nu_{c})-P_{m_{c}}^{c}(\nu_{c}-1) = P_{m_{c}-1}^{c}(\nu_{c}-1),
\nonumber\\
P_{m_{c}-1}^{c}(\nu_{c}-1)-P_{m_{c}-1}^{c}(\nu_{c}-2) = P_{m_{c}-2}^{c}(\nu_{c}-2); 
\nonumber\\
P_{m_{c}}^{c}(\nu_{c}+1)-P_{m_{c}}^{c}(\nu_{c}) = P_{m_{c}-1}^{c}(\nu_{c}); 
\nonumber\\
P_{m_{c}-1}^{c}(\nu_{c}+1)-P_{m_{c}-1}^{c}(\nu_{c}) = P_{m_{c}-2}^{c}(\nu_{c}).  
\label{eq:Ident11}
\end{eqnarray} 

Eq. (\ref{eq:CommRules1}) permits us to bring all raising or lowering $x$ operators to the right hand side to act directly on the Kronecker symbol wavefunction $\psi_{I}=\delta_{\sum_{i}i\nu_{i},n}$ as 
\begin{eqnarray}
ab\nu_{a}(\nu_{b}-\delta_{ab})\widehat{x}_{a}^{-}\widehat{x}_{b}^{-}\widehat{x}_{a+b}^{+}\psi_{I}=ab\nu_{a}(\nu_{b}-\delta_{ab})\psi_{I}
\label{eq:UnitActa}
\end{eqnarray}
or 
\begin{eqnarray}
(a+b)\nu_{a+b}\widehat{x}_{a}^{+}\widehat{x}_{b}^{+}\widehat{x}_{a+b}^{-}\psi_{I}=(a+b)\nu_{a+b}\psi_{I}. 
\label{eq:UnitActb}
\end{eqnarray}
The action of the product of three operators does not change the wavefunction if $\nu_{a}, \nu_{b} \neq 0$ in Eq. (\ref{eq:UnitActa}) or if  
$\nu_{a+b} \neq 0$ in Eq. (\ref{eq:UnitActb}) bringing it to zero otherwise. However in that case (e. g. $\nu_{a}=0$ or $\nu_{b}=0$ in Eq. (\ref{eq:UnitActa})) the operator action results in the zero answer because of the presence of operators $\nu_{a}$, $\nu_{b}$ or $\nu_{a+b}$. Consequently we can skip the product of $x$-operators if they are applied directly to the unit function $\psi_{I}$ and then we can also skip the unit function $\psi_{I}$ to avoid the complexity in the notations. 

The main target of the further mathematical consideration is to reexpress all polynomials bringing their shifted arguments $\nu_{i}\pm 1$ back to  $\nu_{i}$ by means of shifting the parameters $m_{i}$ and then simplifying the like terms that should lead to the target Eq. (\ref{eq:RecEq}). Additional identities related to Laguerre polynomials \cite{Laguerre}  will also be used used for this purpose. These identities are summarized below  
\begin{eqnarray}
\nu_{c}P_{m_{c}-1}^{c}(\nu_{c}-1)=m_{c}P_{m_{c}}^{c}(\nu_{c})+P_{m_{c}-1}^{c}(\nu_{c})/c; 
\nonumber\\
\nu_{c}(\nu_{c}-1)P_{m_{c}-2}^{c}(\nu_{c}-2)=m_{c}(m_{c}-1)P_{m_{c}}^{c}(\nu_{c})+2(m_{c}-1)P_{m_{c}-1}^{c}(\nu_{c})/c+P_{m_{c}-2}^{c}(\nu_{c})/c^2;
\nonumber\\
c\nu_{c}P_{m_{c}-1}^{c}(\nu_{c})=cm_{c}P_{m_{c}}^{c}(\nu_{c})+(cm_{c}+1-c)P_{m_{c}-1}^{c}(\nu_{c})+P_{m_{c}-2}^{c}(\nu_{c});
\nonumber\\
c\nu_{c}P_{m_{c}}^{c}(\nu_{c})=c(m_{c}+1)P_{m_{c}+1}^{c}(\nu_{c})+(cm_{c}+1)P_{m_{c}}^{c}(\nu_{c})+P_{m_{c}-1}^{c}(\nu_{c}).  
\label{eq:Ident11a}
\end{eqnarray} 
Also the identity defining the principal quantum number $n$ Eq. (\ref{eq:Ident1}) will be used.

To evaluate the commutator of the Hamiltonian with the arbitrarily product of  Laguerre polynomials $\psi_{\{m\}}(\{\nu\})$ we consider the most general form of the commutators  with the first (top line) and second (bottom line) parts of the Hamiltonian Eq. (\ref{eq:OrigPr}) enumerated by the numbers (1) for the first part with $a \neq b$, (2) for the first part with $a = b$, (3) for the second part with $a \neq b$, (4) for the second part with $a = b$. Without restricting the generality of the consideration we assume $a>b$ in the cases $a \neq b$. The Laguerre polynomials commuting with the part of the Hamiltonian under consideration are skipped for the sake of simplicity 

One can evaluate the commutators breaking it into more enumerated parts until the desirable form of expressions designated by the ``*" symbol is attained for each part. For the first type of commutators enumerated by the index  (1) we get 
\begin{eqnarray}
(1)=a\nu_{a}b\nu_{b}[x_{a}^{-}x_{b}^{-}x_{a+b}^{+}, P_{m_{a}}^{a}(\nu_{a})P_{m_{b}}^{b}(\nu_{b})P_{m_{a+b}}^{a+b}(\nu_{a+b})]=
\nonumber\\
=a\nu_{a}b\nu_{b}P_{m_{a+b}}^{a+b}(\nu_{a+b})[x_{a}^{-}x_{b}^{-}, P_{m_{a}}^{a}(\nu_{a})P_{m_{b}}^{b}(\nu_{b})] (1.1) +
a\nu_{a}b\nu_{b}P_{m_{a}}^{a}(\nu_{a}-1)P_{m_{b}}^{b}(\nu_{b}-1)[x_{a+b}^{+}, P_{m_{a+b}}^{a+b}(\nu_{a+b})] (1.2). 
\label{eq:CommGen1}
\end{eqnarray}

The first term (1.1) in Eq. (\ref{eq:CommGen1}) can be evaluated as 
\begin{eqnarray}
(1.1)=a\nu_{a}b\nu_{b}P_{m_{a+b}}^{a+b}(\nu_{a+b})[x_{a}^{-}x_{b}^{-}, P_{m_{a}}^{a}(\nu_{a})P_{m_{b}}^{b}(\nu_{b})] =
\nonumber\\
=a\nu_{a}b\nu_{b}P_{m_{a+b}}^{a+b}(\nu_{a+b})P_{m_{a}}^{a}(\nu_{a})[x_{b}^{-}, P_{m_{b}}^{b}(\nu_{b})]+a\nu_{a}b\nu_{b}P_{m_{a+b}}^{a+b}(\nu_{a+b})P_{m_{b}}^{b}(\nu_{b})[x_{a}^{-}, P_{m_{a}}^{a}(\nu_{a})]+
\nonumber\\
+a\nu_{a}b\nu_{b}P_{m_{a+b}}^{a+b}(\nu_{a+b})[x_{a}^{-}, P_{m_{a}}^{a}(\nu_{a})][x_{b}^{-}, P_{m_{b}}^{b}(\nu_{b})]=
\nonumber\\
=-a\nu_{a}b\nu_{b}P_{m_{a+b}}^{a+b}(\nu_{a+b})P_{m_{a}}^{a}(\nu_{a})P_{m_{b}-1}^{b}(\nu_{b}-1)~(1.1.1)-a\nu_{a}b\nu_{b}P_{m_{a+b}}^{a+b}(\nu_{a+b})P_{m_{a}-1}^{a}(\nu_{a}-1)P_{m_{b}}^{b}(\nu_{b})~ (1.1.2)+
\nonumber\\
+a\nu_{a}b\nu_{b}P_{m_{a+b}}^{a+b}(\nu_{a+b})P_{m_{a}-1}^{a}(\nu_{a}-1)P_{m_{b}-1}^{b}(\nu_{b}-1) ~(1.1.3)=
\nonumber\\ 
=-a\nu_{a}P_{m_{a+b}}^{a+b}(\nu_{a+b})P_{m_{a}}^{a}(\nu_{a})bm_{b}P_{m_{b}}^{b}(\nu_{b})~(1.1.1.A*) -a\nu_{a}P_{m_{a+b}}^{a+b}(\nu_{a+b})P_{m_{a}}^{a}(\nu_{a})P_{m_{b}-1}^{b}(\nu_{b})~(1.1.2.A*) -
\nonumber\\
-b\nu_{b}P_{m_{a+b}}^{a+b}(\nu_{a+b})P_{m_{b}}^{b}(\nu_{b})am_{a}P_{m_{a}}^{a}(\nu_{a})(1.1.1.B*)-b\nu_{b}P_{m_{a+b}}^{a+b}(\nu_{a+b})P_{m_{b}}^{b}(\nu_{b})P_{m_{a}-1}^{a}(\nu_{a}) ~ (1.1.2.B*)+
\nonumber\\
+am_{a}bm_{b}P_{m_{a+b}}^{a+b}(\nu_{a+b})P_{m_{a}}^{a}(\nu_{a})P_{m_{b}}^{b}(\nu_{b})~(1.1.3.A*) +am_{a}P_{m_{a+b}}^{a+b}(\nu_{a+b})P_{m_{a}}^{a}(\nu_{a})P_{m_{b}-1}^{b}(\nu_{b})~(1.1.3.B*)+
\nonumber\\
+bm_{b}P_{m_{a+b}}^{a+b}(\nu_{a+b})P_{m_{a}-1}^{a}(\nu_{a})P_{m_{b}}^{b}(\nu_{b})~(1.1.3.C*)+P_{m_{a+b}}^{a+b}(\nu_{a+b})P_{m_{a}-1}^{a}(\nu_{a})P_{m_{b}-1}^{b}(\nu_{b}) ~(1.1.3.D*). 
\label{eq:CommGen3}
\end{eqnarray}
The next stage of evaluation will be performed with respect to the sums over $a$ and $b$ after the first stage calculations  of other terms will be completed.

The second term can be evaluated as 
\begin{eqnarray}
(1.2)= a\nu_{a}b\nu_{b}P_{m_{a}}^{a}(\nu_{a}-1)P_{m_{b}}^{b}(\nu_{b}-1)[x_{a+b}^{+}, P_{m_{a+b}}^{a+b}(\nu_{a+b})]=
\nonumber\\
= P_{m_{a+b}-1}^{a+b}(\nu_{a+b})a(m_{a}+1)b(m_{b}+1)P_{m_{a}+1}^{a}(\nu_{a})P_{m_{b}+1}^{b}(\nu_{b}) ~(1.2.1*)+ 
\nonumber\\ 
+P_{m_{a+b}-1}^{a+b}(\nu_{a+b})a(m_{a}+1)P_{m_{a}+1}^{a}(\nu_{a})P_{m_{b}}^{b}(\nu_{b})~ (1.2.2*) + 
\nonumber\\
+P_{m_{a+b}-1}^{a+b}(\nu_{a+b})b(m_{b}+1)P_{m_{a}}^{a}(\nu_{a})P_{m_{b}+1}^{b}(\nu_{b})~ (1.2.3*) + P_{m_{a+b}-1}^{a+b}(\nu_{a+b})P_{m_{a}}^{a}(\nu_{a})P_{m_{b}}^{b}(\nu_{b}) ~(1.2.4*). 
\label{eq:CommGen2}
\end{eqnarray}
The contribution (1.2.1*) describes the desirable action of the operator product $y_{a}^{-}y_{b}^{-}y_{a+b}^{+}$ term on the set $m$ (cf. Eq. (\ref{eq:RecEq})).

Special consideration should be given to the case $a=b$.  The involved commutator can be expressed as 
\begin{eqnarray}
(2)=a^2 \nu_{a}(\nu_{a}-1)[x_{a}^{-2}x_{2a}^{+}, P_{m_{a}}^{a}(\nu_{a})P_{m_{2a}}^{2a}(\nu_{2a})]=
\nonumber\\
=a^2 \nu_{a}(\nu_{a}-1) P_{m_{2a}}^{2a}(\nu_{2a})[x_{a}^{-2}, P_{m_{a}}^{a}(\nu_{a})] (2.1) +
a^2 \nu_{a}(\nu_{a}-1)P_{m_{a}}^{a}(\nu_{a}-2)[x_{2a}^{+}, P_{m_{2a}}^{2a}(\nu_{2a})] (2.2). 
\label{eq:CommGen1id}
\end{eqnarray}

The first term (2.1) in Eq. (\ref{eq:CommGen1id}) can be expressed as 
\begin{eqnarray}
(2.1)=a^2 \nu_{a}(\nu_{a}-1) P_{m_{2a}}^{2a}(\nu_{2a})[x_{a}^{-2}, P_{m_{a}}^{a}(\nu_{a})]=
\nonumber\\
=P_{m_{2a}}^{2a}(\nu_{2a})a^2 \nu_{a}(\nu_{a}-1)(-2P_{m_{a}-1}^{a}(\nu_{a}-1)+P_{m_{a}-2}^{a}(\nu_{a}-2))=
\nonumber\\
=-2P_{m_{2a}}^{2a}(\nu_{2a})a^2 (\nu_{a}-1)m_{a}P_{m_{a}}^{a}(\nu_{a}) ~ (2.1.1*) -
\nonumber\\
-2P_{m_{2a}}^{2a}(\nu_{2a})a \nu_{a}P_{m_{a}-1}^{a}(\nu_{a}) (2.1.2) + 2P_{m_{2a}}^{2a}(\nu_{2a})a P_{m_{a}-1}^{a}(\nu_{a}) ~(2.1.3*)+
\nonumber\\
+P_{m_{2a}}^{2a}(\nu_{2a})a^2 m_{a}(m_{a}-1)P_{m_{a}}^{a}(\nu_{a})~(2.1.4*)+P_{m_{2a}}^{2a}(\nu_{2a})2a(m_{a}-1)P_{m_{a}-1}^{a}(\nu_{a})~(2.1.5*)+
\nonumber\\
+P_{m_{2a}}^{2a}(\nu_{2a})P_{m_{a}-2}^{a}(\nu_{a})~ (2.1.6*).
\label{eq:CommGen3id}
\end{eqnarray}
The term (2.1.2) needs further evaluation that can be performed using identities Eq. (\ref{eq:Ident11a}) as 
\begin{eqnarray}
(2.1.2)=-2a\nu_{a}P_{m_{2a}}^{2a}(\nu_{2a})P_{m_{a}-1}^{a}(\nu_{a})=
\nonumber\\
-2am_{a}P_{m_{a}}^{a}(\nu_{a})P_{m_{2a}}^{2a}(\nu_{2a}) ~(2.1.2.1*)-
\nonumber\\ 
-2(am_{a}+1-a)P_{m_{a}-1}^{a}(\nu_{a})P_{m_{2a}}^{2a}(\nu_{2a}) ~(2.1.2.2*)-
\nonumber\\ 
-2P_{m_{a}-2}^{a}(\nu_{a})P_{m_{2a}}^{2a}(\nu_{2a}) ~(2.1.2.3*).
\label{eq:212}
\end{eqnarray}

The second term can be evaluated as 
\begin{eqnarray}
(2.2)=a^2 \nu_{a}(\nu_{a}-1)P_{m_{a}}^{a}(\nu_{a}-2)[x_{2a}^{+}, P_{m_{2a}}^{2a}(\nu_{2a})]=
\nonumber\\
=(a^2 (m_{a}+2)(m_{a}+1)P_{m_{a}+2}^{a}(\nu_{a})+2a(m_{a}+1)P_{m_{a}+1}^{a}(\nu_{a})+
\nonumber\\
+P_{m_{a}}^{a}(\nu_{a}))P_{m_{2a}-1}^{2a}(\nu_{2a})=
\nonumber\\
a^2 (m_{a}+2)(m_{a}+1)P_{m_{a}+2}^{a}(\nu_{a})P_{m_{2a}-1}^{2a}(\nu_{2a})~ (2.2.1*)+
\nonumber\\
+2a(m_{a}+1)P_{m_{a}+1}^{a}(\nu_{a})P_{m_{2a}-1}^{2a}(\nu_{2a})~(2.2.2*)+
\nonumber\\
+P_{m_{a}}^{a}(\nu_{a}))P_{m_{2a}-1}^{2a}(\nu_{2a}) ~(2.2.3*);
\label{eq:CommGen2id}
\end{eqnarray}

Two more types of commutators (3) and (4) related to the second part of the Hamiltonian need to be evaluated. For the first type ($a\neq b$) one has
\begin{eqnarray}
(3)=(a+b)\nu_{a+b}[x_{a}^{+}x_{b}^{+}x_{a+b}^{-}, P_{m_{a}}^{a}(\nu_{a})P_{m_{b}}^{b}(\nu_{b})P_{m_{a+b}}^{a+b}(\nu_{a+b})]=
\nonumber\\
= (a+b)\nu_{a+b}P_{m_{a+b}}^{a+b}(\nu_{a+b})[x_{a}^{+}x_{b}^{+}, P_{m_{a}}^{a}(\nu_{a})P_{m_{b}}^{b}(\nu_{b})] ~(3.1)+
\nonumber\\
+(a+b)\nu_{a+b}P_{m_{a}}^{a}(\nu_{a})P_{m_{b}}^{b}(\nu_{b})[x_{a+b}^{-}, P_{m_{a+b}}^{a+b}(\nu_{a+b})] ~(3.2)+
\nonumber\\
+(a+b)\nu_{a+b}[x_{a}^{+}x_{b}^{+}, P_{m_{a}}^{a}(\nu_{a})P_{m_{b}}^{b}(\nu_{b})][x_{a+b}^{-}, P_{m_{a+b}}^{a+b}(\nu_{a+b})] ~(3.3).  
\label{eq:CommGen31}
\end{eqnarray}
Each part can be evaluated separately. Consider them following the order in Eq. (\ref{eq:CommGen31})
\begin{eqnarray}
(3.1)= (a+b)\nu_{a+b}P_{m_{a+b}}^{a+b}(\nu_{a+b})[x_{a}^{+}x_{b}^{+}, P_{m_{a}}^{a}(\nu_{a})P_{m_{b}}^{b}(\nu_{b})] =
\nonumber\\
= (a+b)\nu_{a+b}P_{m_{a+b}}^{a+b}(\nu_{a+b}) P_{m_{a}-1}^{a}(\nu_{a})P_{m_{b}-1}^{b}(\nu_{b}) ~(3.1.1)+
\nonumber\\
+(a+b)\nu_{a+b}P_{m_{a+b}}^{a+b}(\nu_{a+b}) P_{m_{a}}^{a}(\nu_{a})P_{m_{b}-1}^{b}(\nu_{b}) ~(3.1.2*)+
\nonumber\\
+(a+b)\nu_{a+b}P_{m_{a+b}}^{a+b}(\nu_{a+b}) P_{m_{a}-1}^{a}(\nu_{a})P_{m_{b}}^{b}(\nu_{b}) ~(3.1.3*).
\label{eq:CommGen31a}
\end{eqnarray}
Using Eq. (\ref{eq:Ident11a}) one can switch to the desirable set of basis functions for the problematic  term (3.1.1) as 
\begin{eqnarray}
(3.1.1)
= (a+b)\nu_{a+b}P_{m_{a+b}}^{a+b}(\nu_{a+b}) P_{m_{a}-1}^{a}(\nu_{a})P_{m_{b}-1}^{b}(\nu_{b}) =
\nonumber\\
=P_{m_{a}-1}^{a}(\nu_{a})P_{m_{b}-1}^{b}(\nu_{b})(a+b)(m_{a+b}+1)P_{m_{a+b}+1}^{a+b}(\nu_{a+b}) ~(3.1.1.1*)+ 
\nonumber\\ 
+P_{m_{a}-1}^{a}(\nu_{a})P_{m_{b}-1}^{b}(\nu_{b})((a+b)m_{a+b}+1)P_{m_{a+b}}^{a+b}(\nu_{a+b})~(3.1.1.2*)+
\nonumber\\ 
+P_{m_{a}-1}^{a}(\nu_{a})P_{m_{b}-1}^{b}(\nu_{b})P_{m_{a+b}-1}^{a+b}(\nu_{a+b})~(3.1.1.3*). 
\label{eq:CommGen31a1}
\end{eqnarray}

The next contribution (3.2) in Eq. (\ref{eq:CommGen31}) can be evaluated as 
\begin{eqnarray}
(3.2)=(a+b)\nu_{a+b}P_{m_{a}}^{a}(\nu_{a})P_{m_{b}}^{b}(\nu_{b})[x_{a+b}^{-}, P_{m_{a+b}}^{a+b}(\nu_{a+b})]=
\nonumber\\
=-(a+b)\nu_{a+b}P_{m_{a}}^{a}(\nu_{a})P_{m_{b}}^{b}(\nu_{b})P_{m_{a+b}-1}^{a+b}(\nu_{a+b}-1)
=-(a+b)m_{a+b}P_{m_{a}}^{a}(\nu_{a})P_{m_{b}}^{b}(\nu_{b})P_{m_{a+b}}^{a+b}(\nu_{a+b}) ~ (3.2.1*)
-\nonumber\\
-P_{m_{a}}^{a}(\nu_{a})P_{m_{b}}^{b}(\nu_{b})P_{m_{a+b}-1}^{a+b}(\nu_{a+b}) ~ (3.2.2*).  
\label{eq:CommGen322}
\end{eqnarray}
Similarly one can evaluate the remaining contribution (3.3)
\begin{eqnarray}
(3.3)=(a+b)\nu_{a+b}[x_{a}^{+}x_{b}^{+}, P_{m_{a}}^{a}(\nu_{a})P_{m_{b}}^{b}(\nu_{b})][x_{a+b}^{-}, P_{m_{a+b}}^{a+b}(\nu_{a+b})] =
\nonumber\\
=-(a+b)\nu_{a+b}P_{m_{a+b}-1}^{a+b}(\nu_{a+b}-1)(P_{m_{a}}^{a}(\nu_{a})P_{m_{b}-1}^{b}(\nu_{b})+P_{m_{a}-1}^{a}(\nu_{a})P_{m_{b}}^{b}(\nu_{b})+P_{m_{a}-1}^{a}(\nu_{a})P_{m_{b}-1}^{b}(\nu_{b}))=  
\nonumber\\
=-(a+b)m_{a+b}P_{m_{a+b}}^{a+b}(\nu_{a+b})P_{m_{a}}^{a}(\nu_{a})P_{m_{b}-1}^{b}(\nu_{b})~ (3.3.1.1*)-
\nonumber\\
-P_{m_{a+b}-1}^{a+b}(\nu_{a+b})P_{m_{a}}^{a}(\nu_{a})P_{m_{b}-1}^{b}(\nu_{b})~(3.3.1.2*)-
\nonumber\\
-(a+b)m_{a+b}P_{m_{a+b}}^{a+b}(\nu_{a+b})P_{m_{a}-1}^{a}(\nu_{a})P_{m_{b}}^{b}(\nu_{b})~ (3.3.2.1*)-
\nonumber\\
-P_{m_{a+b}-1}^{a+b}(\nu_{a+b})P_{m_{a}-1}^{a}(\nu_{a})P_{m_{b}}^{b}(\nu_{b})~(3.3.2.2*)-
\nonumber\\
-(a+b)m_{a+b}P_{m_{a+b}}^{a+b}(\nu_{a+b})P_{m_{a}-1}^{a}(\nu_{a})P_{m_{b}-1}^{b}(\nu_{b}) (3.3.3.1*)-
\nonumber\\
-P_{m_{a+b}-1}^{a+b}(\nu_{a+b})P_{m_{a}-1}^{a}(\nu_{a})P_{m_{b}-1}^{b}(\nu_{b}) (3.3.3.2*).
\label{eq:CommGen33}
\end{eqnarray}

The remaining commutator (4) is similar to Eq. (\ref{eq:CommGen31}) but describes  the case $a=b$. The related equation takes the form 
\begin{eqnarray}
(4)=2a\nu_{2a}[x_{a}^{+2}x_{2a}^{-}, P_{m_{a}}^{a}(\nu_{a})P_{m_{2a}}^{2a}(\nu_{2a})]=
\nonumber\\
= 2a\nu_{2a}P_{m_{2a}}^{2a}(\nu_{2a})[x_{a}^{+}x_{a}^{+}, P_{m_{a}}^{a}(\nu_{a})] (4.1)+
\nonumber\\
+(2a)\nu_{2a}P_{m_{a}}^{a}(\nu_{a})[x_{2a}^{-}, P_{m_{2a}}^{2a}(\nu_{2a})] (4.2)+
\nonumber\\
+(2a)\nu_{2a}[x_{a}^{+}x_{a}^{+}, P_{m_{a}}^{a}(\nu_{a})][x_{2a}^{-}, P_{m_{2a}}^{2a}(\nu_{2a})] (4.3).  
\label{eq:CommGen41}
\end{eqnarray}
Each part can be evaluated separately. Consider them following the order in Eq. (\ref{eq:CommGen41})
\begin{eqnarray}
(4.1)= 2a\nu_{2a}P_{m_{2a}}^{2a}(\nu_{2a})[x_{a}^{+}x_{a}^{+}, P_{m_{a}}^{a}(\nu_{a})] =
\nonumber\\
= 2a\nu_{2a}P_{m_{2a}}^{2a}(\nu_{2a}) P_{m_{a}-2}^{a}(\nu_{a}) ~ (4.1.1)+
\nonumber\\
+2\cdot 2a\nu_{2a}P_{m_{2a}}^{2a}(\nu_{2a}) P_{m_{a}-1}^{a}(\nu_{a})~ (4.1.2*). 
\label{eq:CommGen41a}
\end{eqnarray}
Using Eq. (\ref{eq:Ident11a}) one can switch to the desirable set of basis functions for the term (4.1.1) in Eq. (\ref{eq:CommGen41a}) as 
\begin{eqnarray}
(4.1.1)
= 2a\nu_{2a}P_{m_{2a}}^{2a}(\nu_{2a}) P_{m_{a}-2}^{a}(\nu_{a}) =
\nonumber\\
=P_{m_{a}-2}^{a}(\nu_{a})2a(m_{2a}+1)P_{m_{2a}+1}^{2a}(\nu_{2a})~ (4.1.1.1*)+ 
\nonumber\\ 
+P_{m_{a}-2}^{a}(\nu_{a})(2am_{2a}+1)P_{m_{2a}}^{2a}(\nu_{2a})~(4.1.1.2*)+
\nonumber\\ 
+P_{m_{a}-2}^{a}(\nu_{a})P_{m_{2a}-1}^{2a}(\nu_{2a})~ (4.1.1.3*). 
\label{eq:CommGen41a1}
\end{eqnarray}
%and correspondingly 
%\begin{eqnarray}
%(4.1.2)= 4a\nu_{2a}P_{m_{2a}}^{2a}(\nu_{2a}) P_{m_{a}-1}^{a}(\nu_{a}) =
%\nonumber\\
%=P_{m_{a}-1}^{a}(\nu_{a})4a(m_{2a}+1)P_{m_{2a}+1}^{2a}(\nu_{2a})~ (4.1.2.1*)+ 
%\nonumber\\ 
%+P_{m_{a}-1}^{a}(\nu_{a})(4am_{2a}+1)P_{m_{2a}}^{2a}(\nu_{2a})~(4.1.2.2*)+
%\nonumber\\ 
%+2P_{m_{a}-1}^{a}(\nu_{a})P_{m_{2a}-1}^{2a}(\nu_{2a}) ~(4.1.2.3*). 
%\label{eq:CommGen41a23}
%\end{eqnarray}

The next contribution (4.2) in Eq. (\ref{eq:CommGen41}) can be evaluated as 
\begin{eqnarray}
(4.2)=2a\nu_{2a}P_{m_{a}}^{a}(\nu_{a})[x_{2a}^{-}, P_{m_{2a}}^{2a}(\nu_{2a})]=
\nonumber\\
=-2a\nu_{2a}P_{m_{a}}^{a}(\nu_{a})P_{m_{2a}-1}^{2a}(\nu_{2a}-1)
=-2am_{2a}P_{m_{a}}^{a}(\nu_{a})P_{m_{2a}}^{2a}(\nu_{2a}) ~(4.2.1*)
-\nonumber\\
-P_{m_{a}}^{a}(\nu_{a})P_{m_{2a}-1}^{2a}(\nu_{2a})~ (4.2.2*).  
\label{eq:CommGen422}
\end{eqnarray}
Similarly one can evaluate the remaining contribution (4.3)
\begin{eqnarray}
(4.3)=2a\nu_{2a}[x_{a}^{+}x_{a}^{+}, P_{m_{a}}^{a}(\nu_{a})][x_{2a}^{-}, P_{m_{2a}}^{2a}(\nu_{2a})] =
\nonumber\\
=-2a\nu_{2a}P_{m_{2a}-1}^{2a}(\nu_{2a}-1)(2P_{m_{a}-1}^{a}(\nu_{a})+P_{m_{a}-2}^{a}(\nu_{a}))=  
\nonumber\\
=-2am_{2a}P_{m_{2a}}^{2a}(\nu_{2a})P_{m_{a}-2}^{a}(\nu_{a})~ (4.3.1.1*)-
\nonumber\\
-P_{m_{2a}-1}^{2a}(\nu_{2a})P_{m_{a}-2}^{a}(\nu_{a}) ~(4.3.1.2*)-
\nonumber\\
-4am_{2a}P_{m_{2a}}^{2a}(\nu_{2a})P_{m_{a}-1}^{a}(\nu_{a}) ~(4.3.2.1*)-
\nonumber\\
-2P_{m_{2a}-1}^{2a}(\nu_{2a})P_{m_{a}-1}^{a}(\nu_{a}) ~(4.3.2.2*).
\label{eq:CommGen43}
\end{eqnarray}

Finally one have to express all results as sums. Unchanged products of Laguerre polynomials are skipped in each sum for the sake of simplicity. We begin with  Eq. (\ref{eq:CommGen2}) and introduce new notations given after each result in its final form designated with the letter $F$. 
\begin{eqnarray}
(1.2.1*)_{sum}= \frac{1}{2}\sum_{b, a\neq b} a(m_{a}+1)b(m_{b}+1)P_{m_{a}+1}^{a}(\nu_{a})P_{m_{b}+1}^{b}(\nu_{b})P_{m_{a+b}-1}^{a+b}(\nu_{a+b}) ~  (F1.1.A);
\nonumber\\ 
(1.2.2+3*)_{sum}=\sum_{b, a\neq b}P_{m_{a+b}-1}^{a+b}(\nu_{a+b})a(m_{a}+1)P_{m_{a}+1}^{a}(\nu_{a})=
\nonumber\\ 
=\sum_{c, a < c}P_{m_{c}-1}^{c}(\nu_{c})a(m_{a}+1)P_{m_{a}+1}^{a}(\nu_{a})~ (F1.1.B)-\sum_{a}P_{m_{2a}-1}^{2a}(\nu_{2a})a(m_{a}+1)P_{m_{a}+1}^{a}(\nu_{a}) ~ (F1.1.B.1); 
\nonumber\\ 
(1.2.4*)_{sum}=\frac{1}{2}\sum_{b, a\neq b}P_{m_{a+b}-1}^{a+b}(\nu_{a+b})=
\nonumber\\
=\frac{1}{2}\sum_{a}(a-1)P_{m_{a}-1}^{a}(\nu_{a})~ (F1.1.C) - \frac{1}{2}\sum_{a}P_{m_{2a}-1}^{2a}(\nu_{2a}) ~ (F1.1.C.1). 
\label{eq:CommGen2S}
\end{eqnarray}

The next group of expressions Eq. (\ref{eq:CommGen3}) can be evaluated as 
\begin{eqnarray}
(1.1.1.A+B*)_{sum}=-\sum_{b, a\neq b}a\nu_{a}bm_{b}=-nm+\sum_{a}a^{2}\nu_{a}m_{a} ~ (F1.2.A);
\nonumber\\
(1.1.2.A+B*)_{sum} =-\sum_{a, b\neq a}b\nu_{b}P_{m_{a}-1}^{a}(\nu_{a})=
\nonumber\\
=-n\sum_{a}P_{m_{a}-1}^{a}(\nu_{a})~ (F1.2.B.1)+
\nonumber\\
+\sum_{a}a\nu_{a}P_{m_{a}-1}^{a}(\nu_{a}) ~ (1.2.B.2); 
\nonumber\\
(1.2.B.2)=\sum_{a}a\nu_{a}P_{m_{a}-1}^{a}(\nu_{a})=(F1.2.B.2.A)+(F1.2.B.2.B)+(F1.2.B.2.C); 
\nonumber\\
(F1.2.B.2.A)=\sum_{a}am_{a}=m ~ (F1.2.B.2.A);
\nonumber\\
(F1.2.B.2.B)=\sum_{a}(am_{a}+1-a)P_{m_{a}-1}^{a}(\nu_{a}) ~ (F1.2.B.2.B);
\nonumber\\
(F1.2.B.2.C)=\sum_{a}P_{m_{a}-2}^{a}(\nu_{a}) ~ (F1.2.B.2.C);
\nonumber\\
(1.1.3.A*)_{sum}=\frac{1}{2}\sum_{b, a\neq b} am_{a}bm_{b}=\frac{m^2}{2} ~(F1.2.C.1) -\frac{1}{2}\sum_{a}a^2m_{a}^2 ~(F1.2.C.2);
\nonumber\\
(1.1.3.B+C*)_{sum}= \sum_{b, a\neq b} am_{a}P_{m_{b}-1}^{b}(\nu_{b})=
\nonumber\\
= m\sum_{a}P_{m_{a}-1}^{a}(\nu_{a})~(F1.3.A) - \sum_{a}am_{a}P_{m_{a}-1}^{a}(\nu_{a})~(F1.3.B); 
\nonumber\\
(1.1.3.D*)_{sum}=\frac{1}{2}\sum_{b, a\neq b}P_{m_{a}-1}^{a}(\nu_{a})P_{m_{b}-1}^{b}(\nu_{b}) ~(F1.4). 
\label{eq:CommGen3S}
\end{eqnarray}

Next we consider the contribution of the case $a=b$ from Eq. (\ref{eq:CommGen1id}) for the first part of the Hamiltonian 
\begin{eqnarray}
(2.2.1*)_{sum}=\frac{1}{2}\sum_{a}a^2 (m_{a}+2)(m_{a}+1)P_{m_{a}+2}^{a}(\nu_{a})P_{m_{2a}-1}^{2a}(\nu_{2a}) ~(F2.1);
\nonumber\\
(2.2.2*)_{sum}=\sum_{a}a(m_{a}+1)P_{m_{a}+1}^{a}(\nu_{a})P_{m_{2a}-1}^{2a}(\nu_{2a})~ (F2.2);
\nonumber\\
(2.2.3*)_{sum}=\frac{1}{2}\sum_{a}P_{m_{a}}^{a}(\nu_{a}))P_{m_{2a}-1}^{2a}(\nu_{2a}) ~(F2.3);
\nonumber\\
(2.1.1*)_{sum}=-\sum_{a}a^2 (\nu_{a}-1)m_{a} ~(F2.4);
\nonumber\\
(2.1.2.1*)_{sum}=-\sum_{a}am_{a}=-m ~ (F2.5.1); 
\nonumber\\ 
(2.1.2.2*)_{sum}=-\sum_{a}(am_{a}+1-a)P_{m_{a}-1}^{a}(\nu_{a}) ~ (F2.5.2);
\nonumber\\ 
(2.1.2.3*)_{sum}=-\sum_{a}P_{m_{a}-2}^{a}(\nu_{a})P_{m_{2a}}^{2a}(\nu_{2a})~ (F2.5.3);
\nonumber\\
(2.1.3*)_{sum}=\sum_{a}a P_{m_{a}-1}^{a}(\nu_{a})~ (F2.6);
\nonumber\\
(2.1.4*)_{sum}= \frac{1}{2}\sum_{a}a^2 m_{a}(m_{a}-1)~ (F2.7);
\nonumber\\
(2.1.5*)_{sum}=\frac{1}{2}\sum_{a}2a(m_{a}-1)P_{m_{a}-1}^{a}(\nu_{a})~ (F2.8);
\nonumber\\
(2.1.6*)_{sum}=\frac{1}{2}\sum_{a}P_{m_{2a}}^{2a}(\nu_{2a})P_{m_{a}-2}^{a}(\nu_{a}) ~ (F2.9);   
\label{eq:CommGen2idS}
\end{eqnarray}

The next contribution is associated with the remaining part of the Hamiltonian ($a\neq b$) Eq. (\ref{eq:CommGen31})  
\begin{eqnarray}
(3.1.1.1*)_{sum}=\frac{1}{2}\sum_{b, a \neq b} P_{m_{a}-1}^{a}(\nu_{a})P_{m_{b}-1}^{b}(\nu_{b})(a+b)(m_{a+b}+1)P_{m_{a+b}+1}^{a+b}(\nu_{a+b}) ~(F3.1);
\nonumber\\
(3.1.1.2*)_{sum}=\frac{1}{2}\sum_{b, a \neq b} P_{m_{a}-1}^{a}(\nu_{a})P_{m_{b}-1}^{b}(\nu_{b})((a+b)m_{a+b}+1) ~(F3.2.1);
\nonumber\\ 
(3.1.1.3*)_{sum}=\frac{1}{2}\sum_{b, a \neq b}P_{m_{a}-1}^{a}(\nu_{a})P_{m_{b}-1}^{b}(\nu_{b})P_{m_{a+b}-1}^{a+b}(\nu_{a+b}) ~(F3.2.2); 
\nonumber\\
(3.1.2+3*)_{sum}
= \sum_{a, b\neq a}(a+b)\nu_{a+b}P_{m_{b}-1}^{b}(\nu_{b}) =
\nonumber\\
=\sum_{a}nP_{m_{a}-1}^{a}(\nu_{a}) ~(F3.2.3.A) - \sum_{a}2a\nu_{2a}P_{m_{a}-1}^{a}(\nu_{a})~(F3.2.3.B)-
\nonumber\\
-\sum_{a}a\nu_{a}P_{m_{a}-1}^{a}(\nu_{a})~(3.2.3.C)-\sum_{a, c<a}P_{m_{a}-1}^{a}(\nu_{a})c\nu_{c}P_{m_{c}}^{c}(\nu_{c}) ~(3.2.3.D);
\nonumber\\
(3.2.3.C)=(F3.2.3.C.1)+(F3.2.3.C.2)+(F3.2.3.C.3);
\nonumber\\
(F3.2.3.C.1)=-\sum_{a}am_{a}=-m ~ (F3.2.3.C.1);
\nonumber\\
(F3.2.3.C.2)=-\sum_{a}(am_{a}+1-a)P_{m_{a}-1}^{a}(\nu_{a})~ (F3.2.3.C.2);
\nonumber\\
(F3.2.3.C.3)=-\sum_{a}P_{m_{a}-2}^{a}(\nu_{a})~ (F3.2.3.C.3);
\nonumber\\
(3.2.3.D)=-\sum_{a, c<a}P_{m_{a}-1}^{a}(\nu_{a})c\nu_{c}P_{m_{c}}^{c}(\nu_{c})=
\nonumber\\
-\sum_{a, c<a}c(m_{c}+1)P_{m_{c}+1}^{c}(\nu_{c})P_{m_{a}-1}^{a}(\nu_{a})~(F3.2.3.D.1)-
\nonumber\\
-\sum_{a, c<a}(cm_{c}+1)P_{m_{c}}^{c}(\nu_{c})P_{m_{a}-1}^{a}(\nu_{a})~(F3.2.3.D.2)-
\nonumber\\
-\sum_{a, c<a}P_{m_{c}-1}^{c}(\nu_{c})P_{m_{a}-1}^{a}(\nu_{a})~ (F3.2.3.D.3);
\nonumber\\
(3.2.1*)_{sum}=-\frac{1}{2}\sum_{b, a\neq b}(a+b)m_{a+b}=-\frac{1}{2}\sum_{a}am_{a}(a-1)~(F3.2.4)+\frac{1}{2}\sum_{a}2am_{2a}~ (F3.2.4.A);  
\nonumber\\
(3.2.2*)_{sum}=-\frac{1}{2}\sum_{b, a\neq b}P_{m_{a+b}-1}^{a+b}(\nu_{a+b})=
\nonumber\\
=-\frac{1}{2}\sum_{a}(a-1)P_{m_{a}-1}^{a}(\nu_{a})~(F3.2.5)+\frac{1}{2}\sum_{a}P_{m_{2a}-1}^{2a}(\nu_{2a}) ~(F3.2.5.A);
\nonumber\\
(3.3.1.1*)_{sum}+(3.3.2.1*)_{sum}=-\sum_{b, a> b}am_{a}P_{m_{b}-1}^{b}(\nu_{b}) ~(F3.2.6)+
\nonumber\\
+\sum_{a}2am_{2a}P_{m_{a}-1}^{a}(\nu_{a}) ~(F3.2.6.A);
\nonumber\\
(3.3.1.2*)_{sum}+(3.3.2.2*)_{sum}=-\sum_{b, a\neq b}P_{m_{a+b}-1}^{a+b}(\nu_{a+b})P_{m_{b}-1}^{b}(\nu_{b})=
\nonumber\\
=-\frac{1}{2}\sum_{b, a\neq b}P_{m_{a}-1}^{a}(\nu_{a})P_{m_{b}-1}^{b}(\nu_{b})~ (F3.2.7) +
\nonumber\\
+\sum_{a}P_{m_{2a}-1}^{2a}(\nu_{2a})P_{m_{a}-1}^{a}(\nu_{a})~ (F3.2.7.A);
\nonumber\\
(3.3.3.1*)_{sum}=-\frac{1}{2}\sum_{b, a\neq b}(a+b)m_{a+b}P_{m_{a}-1}^{a}(\nu_{a})P_{m_{b}-1}^{b}(\nu_{b}) ~ (F3.2.8);
\nonumber\\
(3.3.3.2*)_{sum}=-\frac{1}{2}\sum_{b, a\neq b}P_{m_{a+b}-1}^{a+b}(\nu_{a+b})P_{m_{a}-1}^{a}(\nu_{a})P_{m_{b}-1}^{b}(\nu_{b})~ (F3.2.9).  
\label{eq:CommGen31aS}
\end{eqnarray}

Finally we consider the second part of the Hamiltonian for $a=b$. The contributions from this part can be expressed as
\begin{eqnarray}
(4.1.1.1*)_{sum}=\frac{1}{2}\sum_{a}P_{m_{a}-2}^{a}(\nu_{a})2a(m_{2a}+1)P_{m_{2a}+1}^{2a}(\nu_{2a})~ (F4.1); 
\nonumber\\ 
(4.1.1.2*)_{sum}=\frac{1}{2}\sum_{a}P_{m_{a}-2}^{a}(\nu_{a})(2am_{2a}+1)P_{m_{2a}}^{2a}(\nu_{2a})~ (F4.2.A);
\nonumber\\ 
(4.1.1.3*)_{sum}=\frac{1}{2}\sum_{a}P_{m_{a}-2}^{a}(\nu_{a})P_{m_{2a}-1}^{2a}(\nu_{2a}) ~(F4.2.B);  
\nonumber\\
(4.1.2*)_{sum}= \frac{1}{2}\sum_{a}4a\nu_{2a}P_{m_{2a}}^{2a}(\nu_{2a}) P_{m_{a}-1}^{a}(\nu_{a}) ~(F4.3).% (- (3.2.3.B) );
\nonumber\\
(4.2.1*)_{sum}=-\frac{1}{2}\sum_{a}2am_{2a} ~ (F4.4.A);
\nonumber\\
(4.2.2*)_{sum}=-\frac{1}{2}\sum_{a}P_{m_{2a}-1}^{2a}(\nu_{2a})~ (F4.4.B);
\nonumber\\
(4.3.1.1*)_{sum}=-\frac{1}{2}\sum_{a}2am_{2a}P_{m_{a}-2}^{a}(\nu_{a})~ (F4.5.A);
\nonumber\\
(4.3.1.2*)_{sum}=-\frac{1}{2}\sum_{a}P_{m_{2a}-1}^{2a}(\nu_{2a})P_{m_{a}-2}^{a}(\nu_{a})~ (F4.5.B);
\nonumber\\
(4.3.2.1*)_{sum}=-\frac{1}{2}\sum_{a}4am_{2a}P_{m_{a}-1}^{a}(\nu_{a})~ (F4.5.C);
\nonumber\\
(4.3.2.2*)_{sum}=-\frac{1}{2}\sum_{a}2P_{m_{2a}-1}^{2a}(\nu_{2a})P_{m_{a}-1}^{a}(\nu_{a})~ (F4.5.D).  
\label{eq:CommGen41aS}
\end{eqnarray}

Finally we collect all contributions with like terms. We begin with the terms corresponding to the same or modified sequences $\{m\}$ representing the partitions of the same number $m$ (e. g. (F4.1) or (F4.4.A)), while the terms representing the partitions of different numbers (e. g. (F4.5.A)) should cancel each other. Indeed, the contribution (F4.1) is represented by the sequences with modified  parameters $m_{a}\rightarrow m_{a}-2, ~ m_{2a}\rightarrow m_{2a}+1$  conserving the sum $m=\sum_{a}am_{a}$, the contribution (F4.4.A) is represented by the same sequence (diagonal term), while the contribution (F4.5.A) is represented by the sequences modified as $m_{a} \rightarrow m_{a}-2$ representing the partition of the different number $m-2a$.  

First we collect all diagonal terms entering with the same basis function $\psi_{\{m\}}(\{\nu\})$. These terms are enumerated by indices (F1.2.A), (F1.2.B.2.A), (F1.2.C.1), (F1.2.C.2), (F2.4), (F2.5.1), (F2.7), (F3.2.3.C.1), (F3.2.4), (F3.2.4.A), (F4.4.A).    
The sum of all terms can be evaluated as 
\begin{eqnarray}
-nm+\sum_{a}a^2\nu_{a}m_{a}~ (F1.2.A) + m ~(F1.2.B.2.A) + \frac{m^2}{2} ~(F1.2.C.1) - \frac{1}{2} \sum_{a}a^2 m_{a}^2  ~(F1.2.C.2)-
\nonumber\\ 
-\sum_{a}a^2m_{a}(\nu_{a}-1)~ (F2.4) -m ~(F2.5.1) + \frac{1}{2}\sum_{a}a^2m_{a}(m_{a}-1) ~(F2.7) -m ~(F3.2.3.C.1)  -
\nonumber\\
-\frac{1}{2}\sum_{a}am_{a}(a-1) ~(F3.2.4) + \frac{1}{2}\sum_{a}2am_{2a} ~(F3.2.4.A)    -\frac{1}{2}\sum_{a}2am_{2a}  ~(F4.4.A) = -nm+\frac{m(m-1)}{2}.
\label{eq:SummDiag}
\end{eqnarray}
This expression taken together with the first term $n(n-1)/2$ in Eq. (\ref{eq:CommNeed}) coincide with the diagonal term in  Eq. (\ref{eq:RecEq}). 

Next we collect the off-diagonal terms represented by the partitions of the same number $m$. They include contributions (F1.1.A), (F2.1), (F3.1)  and (F4.1) which can be written as 
\begin{eqnarray}
I_{++-}^{a\neq b}=\frac{1}{2}\sum_{b, a\neq b} a(m_{a}+1)b(m_{b}+1)P_{m_{a}+1}^{a}(\nu_{a})P_{m_{b}+1}^{b}(\nu_{b})P_{m_{a+b}-1}^{a+b}(\nu_{a+b}) ~(F1.1.A);
\nonumber\\ 
I_{++-}^{a=b}=\frac{1}{2}\sum_{a}a^2 (m_{a}+2)(m_{a}+1)P_{m_{a}+2}^{a}(\nu_{a})P_{m_{2a}-1}^{2a}(\nu_{2a}) ~(F2.1);
\nonumber\\
I_{+--}^{a\neq b}=\frac{1}{2}\sum_{b, a \neq b} P_{m_{a}-1}^{a}(\nu_{a})P_{m_{b}-1}^{b}(\nu_{b})(a+b)(m_{a+b}+1)P_{m_{a+b}+1}^{a+b}(\nu_{a+b}) ~(F3.1);
\nonumber\\
I_{+--}^{a=b}=\frac{1}{2}\sum_{a}P_{m_{a}-2}^{a}(\nu_{a})2a(m_{2a}+1)P_{m_{2a}+1}^{2a}(\nu_{2a}) ~(F4.1).
\label{eq:SummOffDiagReal}
\end{eqnarray}  
The notations reflect the related anharmonic interaction actions for processes $(a, b) \leftarrow\rightarrow (a+b)$ in separately considered cases $(a\neq b)$ or $a=b$. 

The contributions Eqs. (\ref{eq:SummDiag}) and  (\ref{eq:SummOffDiagReal}) added together are equivalent to the recursive equation Eq. (\ref{eq:RecEq}). It is obvious for the diagonal term.  For off-diagonal terms it becomes clear if we consider the total expression for the  wavefunction amplitude Eq. (\ref{eq:SolAnz}) which will appear in the right hand side of Eq. (\ref{eq:RecEq}) with several polynomial products modified as in Eq. (\ref{eq:SummOffDiagReal}). For instance the contribution $(F.1.1.A)$ can be expressed in the form 
\begin{eqnarray}
\frac{1}{2}\sum_{b, a\neq b}\left(\sum_{\{m\}}d_{\{m\}}y_{a}^{+}y_{b}^{+}y_{a+b}^{-}\left[m_{a}m_{b}\Psi_{\{m\}}(\{\nu\})\right]\psi_{I}\right), ~ \psi_{I}=\delta_{\sum_{i}im_{i},m}.
\label{eq:SummOffDiagRealIll}
\end{eqnarray}  
Since the action  of the operator product $y_{a}^{-}y_{b}^{-}y_{a+b}^{+}$ to the expression in parenthesis Eq. (\ref{eq:SummOffDiagRealIll}) does not change this expression because it shifts the summation indices without modifying the whole sum (the ``border" terms with $m_{a}=0$ or $m_{b}=0$ do not contribute because of the factor $m_{a}m_{b}$) one can rewrite this sum shifting the summation indices in the internal sum as $m_{a}\rightarrow m_{a}+1$, $m_{b}\rightarrow m_{b}+1$, $m_{a+b}\rightarrow m_{a+b}-1$ in   the form 
\begin{eqnarray}
\left(\sum_{\{m\}}\Psi_{\{m\}}(\{\nu\})\frac{1}{2}\sum_{b, a\neq b}m_{a}m_{b}y_{a}^{-}y_{b}^{-}y_{a+b}^{+}d_{\{m\}}\psi_{I}\right). 
\label{eq:SummOffDiagRealIll1}
\end{eqnarray} 
The expression multiplied by the basis function $\Psi_{\{m\}}(\{\nu\})$ is identical to the first term in the right hand side of Eq. (\ref{eq:RecEq}) for different indices $a$ and $b$. Similar transformations can be applied to the other three terms in Eq. (\ref{eq:SummOffDiagReal}) reproducing all other contributions in Eq. (\ref{eq:RecEq}) which can be then obtained setting the coefficients with the like terms $\Psi_{\{m\}}(\{\nu\})$ equal to zero, provided that all other off-diagonal terms corresponding to the partitions of a numbers different from $m$ are compensated with each other. One should notice that although the solution for the amplitude $d_{\{m\}}$ satisfying Eq. (\ref{eq:RecEq}) if nontrivial is definitely the eigenstate of the problem other solutions can exist because the basis functions $\Psi_{\{m\}}(\{\nu\})$ are not necessarily linearly independent. For instance for $n=3$, $m=2$ the basis functions $\psi_{(2, 0)}=P_{2}^{1}(\nu_{1})=\nu_{1}(\nu_{1}-1)/2-\nu_{1}+1/2$ and $\psi_{(0, 1)}=P_{1}^{2}(\nu_{2})=\nu_{2}-1/2$  are related to each other as $\psi_{(2, 0)}=-\psi_{(0, 1)}$ for all integer partitions of $n=3$ including $(\nu_{1}, \nu_{2}, \nu_{3})=(3, 0, 0)$, $(1, 1, 0)$, $(0, 0, 1)$. This nonorthogonality is a source of trivial solutions. 
However, according to numerical study all eigenstates probed yet for $n\leq 25$ can be represented by integer number sequences as described in the main body of the manuscript.

Consider the remaining off-diagonal terms, which indeed cancel each other as one can see below. 
First we can perform some obvious cancellations including $(F1.2.B.2.B)+(F2.5.2)=0$, $(F1.2.B.1)+(F3.2.3.A)=0$, $(F3.2.2)+(F3.2.9)=0$, $(F1.2.B.2.C)+(F3.2.3.C.3)=0$, $(F1.1.C)+(F3.2.5)=0$, $(F1.1.C.1)+(F3.2.5.A)=0$, $(F1.1.B.1)+(F2.2)=0$, $(F1.4)+(F3.2.1)+(F3.2.3.D.3)+(F3.2.7)+(F3.2.8)=0$, $(F4.2.B)+(F4.5.B)=0$, 
$(F3.2.6.A)+(F4.5.C)=0$, $(F4.3)+(F3.2.3.B)=0$, $(F1.1.B)+(F3.2.3.D.1)=0$, $(F2.3)+(F4.4.B)=0$, $(F2.5.3)+(F2.9)+(F4.2.A)+(F4.5.A)=0$, $(F1.3.B)+(F2.6)+(F2.8)=0$, $(F3.2.7.A)+(F4.5.D)=0$. The sum of remaining terms is also equal to zero as shown below %(both expressions in square brackets are obviously zero)
\begin{eqnarray} 
 m\sum_{a}P_{m_{a}-1}^{a}(\nu_{a}) ~ (F1.3.A) 
-\sum_{a}(am_{a}+1-a)P_{m_{a}-1}^{a}(\nu_{a}) ~ (F3.2.3.C.2)- 
\nonumber\\
-\sum_{c, a<c}(am_{a}+1)P_{m_{a}}^{a}(\nu_{a})P_{m_{c}-1}^{c}(\nu_{c}) ~ (F3.2.3.D.2)-
\sum_{c, a<c}cm_{c}P_{m_{a}-1}^{a}(\nu_{a}) ~ (F3.2.6)=
\nonumber\\
=\sum_{a}P_{m_{a}-1}^{a}(\nu_{a})\left[m-\sum_{c: a<c}cm_{c}-\sum_{c: a>c}cm_{c}-am_{a}\right]+
\nonumber\\
+\sum_{a}P_{m_{a}-1}^{a}\left[-am_{a}+am_{a}+\sum_{c:c<a}1-(a-1)\right]=0.  
\label{eq:Remainder}
\end{eqnarray}
 Thus the validity of Eq. (\ref{eq:RecEq}) and, correspondingly, of Eq. (\ref{eq:SolAnz}) is proved.

\section{Generating function for wavefunction amplitudes}

Since  eigenfunction amplitudes for a certain principal quantum number $n$ can be expressed through the eigenfunctions of the same problem with different principal quantum number $m$ Eq. (\ref{eq:SolAnz}) one can continue this procedure adding more numbers until the last quantum number approaches zero.  Then possible eigenfunctions of the problem with the principal quantum number $n$ can be described by a  sequence of numbers $k_{0}=n, k_{1}, k_{2},...k_{p-1}, k_{p}=0$. The wavefunction $c_{\{\nu\}}$ for a certain integer partition $\{\nu\}$ of number $n$ representing the resonant sequence of population numbers satisfying the conservation law Eq. (\ref{eq:Ident1})  can be expanded over the basis of the products of Laguerre polynomials $\psi_{\{m\}}(\{\nu\})$ as 
\begin{eqnarray}
c_{\{\nu\}}=\sum_{\{m\}}\psi_{\{m\}}(\{\nu\})d_{\{m\}},
\label{eq:wfrecurs1}
\end{eqnarray}
where the sequences $\{m\}$ represent all integer partitions of the number $k_{1}$. This procedure can be written in the matrix form 
\begin{eqnarray}
{\bf c}=\widehat{M}^{n, k_{1}}{\bf d},
\label{eq:wfrecursMatr}
\end{eqnarray}
where ``vectors" ${\bf c}$ and ${\bf d}$ represent wavefunction amplitudes in the partition spaces for numbers $n$ and $k_{1}$ ($m$) and the matrix element between partitions of numbers $n$ and $k_{1}$ is given by the corresponding product of Laguerre polynomials Eq. (\ref{eq:LaguerreProdApp}).  Remember that if $k_{1}=0$ one can set all amplitudes $d_{\{m\}}=1$. 

Repeating the procedure Eq. (\ref{eq:wfrecurs1}) until reaching the last non-zero number in the sequence $\{k\}$ one can express the eigenfunctions in terms of the product of matrices $\widehat{M}$ by the vector of all unities ${\bf I}_{k_{p-1}}$ of the size equal to the number of integer partitions of the number $k_{p-1}$ as 
\begin{eqnarray}
{\bf c}=\left[\prod_{i=1}^{p-2}\widehat{M}^{k_{i}, k_{i+1}}\right]\mathbf{I}_{k_{p-1}}.
\label{eq:wfrecursMatr1}
\end{eqnarray}

This expression is complicated and we cannot evaluate it in general. Yet it is possible to calculate the related generating function defined as
\begin{eqnarray}
G(\{\nu\}, \{y\})=\sum_{\{m\}_{1}}y_{1}^{\eta(\{m\}_{1})}\psi_{\{m\}_{1}}(\{\nu\})\sum_{\{m\}_{2}}y_{2}^{\eta(\{m\}_{2})}\psi_{\{m\}_{2}}(\{m\}_{1})...
\sum_{\{m\}_{p-1}}y_{p-1}^{\eta(\{m\}_{p-1})},
\label{eq:wfderfun}
\end{eqnarray}
where $\eta(\{m\})=\sum_{i}im_{i}$ is a number whose partition is realized by a sequence $\{m\}$. The wavefunction amplitudes for the specific partition $\{\nu\}$ and sequence $\{k\}$ can be found  as the coefficients of the generating function Eq. (\ref{eq:wfderfun}) polynomial expansion with the products $\prod_{i=1}^{p-1}y_{i}^{k_{i}}$. 

The generating function can be calculated using the identity for Laguerre polynomials which reads \cite{Gradshtein}
\begin{eqnarray}
\sum_{m=0}^{\infty}y^{m}L_{m}^{(n-m)}(x)=e^{-xy}(1+y)^{n}. 
\label{eq:LagIdent}
\end{eqnarray}
Using this identity one can evaluate the generating function Eq. (\ref{eq:wfderfun}). First this function can be separated into the products of contributions of Laguerre polynomials of different arguments $1/j$ where $j$ stands for the $j^{th}$ phonon state.  Then the most right sum in Eq. (\ref{eq:wfderfun}) for the specific $j$ can be evaluated using Eq. (\ref{eq:LagIdent}) as 
\begin{eqnarray}
S_{p-1}(j)=\exp\left(-\frac{y_{p-1}^{j}}{j}\right)(1+y_{p-1}^{j})^{m(p-2, j)},  
\label{eq:LagIdentAppl1}
\end{eqnarray}
where $m(p-2, j)$ is the value of $k^{th}$ number in the sequence $\{m\}_{p-2}$. The next to the most right  summation in Eq. (\ref{eq:wfderfun})  yields 
\begin{eqnarray}
S_{p-2}(j)=\exp\left(-\frac{y_{p-2}^{j}}{j}-\frac{y_{p-2}^{j}y_{p-1}^{j}}{j}-\frac{y_{p-1}^{j}}{j}\right)(1+y_{p-2}^{j}+y_{p-2}^{j}y_{p-1}^{j})^{m(p-3, j)}.   
\label{eq:LagIdentAppl2}
\end{eqnarray}
Repeating this procedure $p-1$ times we can express the contribution of the specific state $j$ as
\begin{eqnarray}
S(j)=\exp\left(-\frac{1}{j}\sum_{f, g (1<f\leq g<p)}\prod_{i=f}^{g}y_{i}^{j}\right)\left[1+y_{1}^{j}+(y_{1}y_{2})^j+...(y_{1}..y_{p-1})^j\right]^{\nu_{j}},    
\label{eq:LagIdentApplk}
\end{eqnarray}
where $\nu_{j}$ is the population number of the $j^{th}$ photon in the state (partition) $\{\nu\}$ of interest. 

Finally taking the product of contributions $S_{j}$ over all states $j=1,2..$ Eq. (\ref{eq:LagIdentApplk}) and using the identity $\sum_{j=1}^{\infty}x^{j}/j=-\ln(1-x)$ we obtain the generating function in the final form   
\begin{eqnarray}
G(\{\nu\}, \{y\})=\prod_{f, g (1<f\leq g<p)}\left[1-\prod_{i=f}^{g}y_{i}\right]\prod_{j=1}^{n}\left[1+y_{1}^{j}+(y_{1}y_{2})^j+...(y_{1}..y_{p-1})^j\right]^{\nu_{j}}  
\label{eq:wfderfunAnsa}
\end{eqnarray}
used in the main body of the manuscript.

Eq. (\ref{eq:wfderfunAnsa}) is too complicated for general analysis of the problem eigenstates. Yet it is sufficient to fully characterize the eigenstates containing non-zero contribution of  the single phonon state $|n>$ represented by the population number sequence $\nu_{k}=\delta_{kn}$. We restrict the consideration to strictly decreasing sequences $\{k\}$ just because we know from the solution that all eigenstates  of interest can be found using these sequences. Then for the state $\nu_{k}=\delta_{kn}$ one can leave only unity from the second product in the right hand side of Eq. (\ref{eq:wfderfunAnsa}) because  otherwise the power of some of the variables $y_{i}$ will approach or exceed $n$ which is not acceptable since these powers should form strictly decreasing sequence $n>k_{1}>k_{2}>k_{3}...>k_{p}=0$. The only acceptable choice of the contribution from  the first product satisfying the sequence selection  requirements can be made taking the composite product of $y_{1}\times (y_{1}y_{2})\times (y_{1}y_{2}y_{3})...\times (y_{1}y_{2}...y_{p-1})$. The related sequence is given by $(n, p-1, p-2, p-3,...0)$ and it defines the eigenstate $\Psi(p)$. There are $n$ such sequences determined by the integer number $p$ changing from $1$ to $n$.  In all cases the absolute value of the amplitude of the wavefunction in the single phonon state is equal unity.

To illustrate the results of the previous paragraph consider the case of $p=3$. Then the part of the generating function that is of interest is given by 
 \begin{eqnarray}
P(\{\nu\}, y_{1}, y_{2})=\prod_{f, g (1<f\leq g<p)}\left[1-\prod_{i=f}^{g}y_{i}\right]=
\nonumber\\
=(1-y_{1})(1-y_{2})(1-y_{1}y_{2})=1-y_{1}-y_{2}-y_{1}y_{2}+y_{1}y_{2}+y_{1}^2 y_{2}+y_{1}y_{2}^2-y_{1}^2y_{2}^2.  
\label{eq:wfderfunAnsEx}
\end{eqnarray}
It is clear that the only contribution of interest is associated with the term $y_{1}^2 y_{2}$ since the power of $y_{1}$ should be larger than the power of $y_{2}$ and they should be both strictly positive. This is in  a full accord with the previous conclusion. 

One can prove using the generating function methods (cf. Eq. (\ref{eq:Genn0})) that the wavefunctions $\Psi(p)$ are normalized by $1$ (see Sec. \ref{sec:Normp}). Using the normalization factor as in Eq. (\ref{eq:n0norm}) we find the normalized by $1$ wavefunction amplitude for the state $|n>$ to be $c_{n}=1/\sqrt{n}$. Since we found $n$ states containing the single phonon state and $n|c_{n}|^2=1$ this gives an additional evidence than our choice is complete. This result is used within the main body of the manuscript to investigate the time evolution of the probability for the system to be in a single phonon state $|n>$.

\section{Normalization of the wavefunction $\Psi(p)$}
\label{sec:Normp}

The proof is separated into two parts. In the first part  we prove that the longest possible strictly decreasing sequence $(n, n-1, ... 0)$ results in the wave function $\Psi(n)$ with amplitudes defined as 
\begin{eqnarray}
c_{\{\nu\}}=(-1)^{\sum_{k}\nu_{k}}(-1)^{n}, ~ n=\sum_{k=1}^{n}k\nu_{k}. 
\label{eq:inv1}
\end{eqnarray}  
The normalization of this wavefunction by $1$ is proved in Sec. \ref{sec:Normn0}  because the eigenstate considered there differs from Eq. (\ref{eq:inv1}) by signs only. In the second part the proof of the normalization by one  will be given for the function $\Psi(p)$ defined by the general sequence $(n, p-1,p-2, ...0)$ ($p\leq n$). 

To prove the first statement assume that it is valid for $p=n-1$ and consider its generalization for $p=n$. Then, according to Eq. (\ref{eq:wfrecurs1}) we can express the wavefunction amplitude for the partitions $\{\nu\}$ of the number $n$ in terms of the wavefunction amplitudes for the partitions $\{m\}$ of the number $n-1$ as 
\begin{eqnarray}
c_{\{\nu\}}=(-1)^{n-1}\sum'_{\{m\}}\prod_{k}L_{m_{k}}^{(\nu_{k}-m_{k})}(1/k)(-1)^{m_{k}}. 
\label{eq:recursp}
\end{eqnarray}
The sum $\sum'$ is taken only over the sequences $\{m\}$ representing the partitions of $n-1$. This expression can be evaluated  using the  generating function $M(\{\nu\}, x)$ defined as 
\begin{eqnarray}
M(\{\nu\}, x)=(-1)^{n-1}\sum_{\{m\}}\prod_{k}L_{m_{k}}^{(\nu_{k}-m_{k})}(1/k)(-1)^{m_{k}}x^{km_{k}}, 
\label{eq:Genrecursp}
\end{eqnarray}
where the sum is taken over all sequences $\{m\}$ of nonnegative population numbers. The wavefunction amplitude of interest is given by the 
expansion coefficient of the function $M(\{\nu\}, x)$ with the factor $x^{n-1}$. 

One can express the generating function Eq. (\ref{eq:recursp}) as a product of independent sums for all integer $k$'s and then evaluate each sum using Eq. (\ref{eq:LagIdent}). Then we get 
\begin{eqnarray}
M(\{\nu\}, x)=\frac{(-1)^{n-1}}{1-x}\prod_{k}\left(1-x^k\right)^{\nu_{k}}.  
\label{eq:Genrecursp1}
\end{eqnarray}
Assume that $l$ is the minimum index $k$ in the product in Eq. (\ref{eq:Genrecursp1}) corresponding to the non-zero population number $\nu_{l}$. Then using the algebraic identity $(1-x^l)/(1-x)=\sum_{q=0}^{l-1}x^{q}$ one can represent Eq. (\ref{eq:Genrecursp1}) as 
\begin{eqnarray}
M(\{\nu\}, x)=(-1)^{n-1}\sum_{q=0}^{l-1}x^{q}\left(1-x^l\right)^{\nu_{l}-1}\prod_{k=l+1}^{n}\left(1-x^k\right)^{\nu_{k}}.  
\label{eq:Genrecursp2}
\end{eqnarray}
The term of interest with the power $x^{n-1}$ is given by the highest power of $x$ term in Eq. (\ref{eq:Genrecursp2}) which reads 
 \begin{eqnarray}
(-1)^{n}x^{n-1}(-1)^{\sum_{k}\nu_{k}}.   
\label{eq:Genrecursp3}
\end{eqnarray}
The coefficient with this term has the form  of Eq. (\ref{eq:inv1}), which proves the first statement. 

Consider the proof of the second statement.  According to Eqs. (\ref{eq:wfrecurs1}) and (\ref{eq:inv1})  the amplitudes of the wavefunction $\Psi(p)$ can be expressed as 
\begin{eqnarray}
c_{\{\nu\}}=(-1)^{n-1}\sum'_{\{m\}}\prod_{k}L_{m_{k}}^{(\nu_{k}-m_{k})}(1/k)(-1)^{m_{k}}, 
\label{eq:StateofInterest}
\end{eqnarray}
where the summation ($\sum'$) is taking over integer partitions of a number $p$ ($\sum_{k}km_{k}=p$).% and  $\sum_{k}k\nu_{k}=n$. 
The normalization of the corresponding wavefunction is given by (cf. Eq. (\ref{eq:n0norm}))
\begin{eqnarray}
\sum'_{\{\nu\}}\frac{|c_{\{\nu\}}|^2}{\prod_{i=1}^{n}i^{\nu_{i}}\nu_{i}!}%=
%\nonumber\\
=\sum'_{\{\nu\}}\frac{1}{\prod_{i=1}^{n}i^{\nu_{i}}\nu_{i}!}\sum'_{\{m\}}\left(\prod_{k}L_{m_{k}}^{(\nu_{k}-m_{k})}(1/k)(-1)^{m_{k}}\right)\sum'_{\{r\}}\left(\prod_{k}L_{r_{k}}^{(\nu_{k}-r_{k})}(1/k)(-1)^{r_{k}}\right),  
\label{eq:pnorm}
\end{eqnarray}
where the summations is over partitions of number $n$ ($\sum_{k}k\nu_{k}=n$) and number $p$ ($\sum_{k}km_{k}=p$, $\sum_{k}kr_{k}=p$). 

As previously the summations over partitions in Eq. (\ref{eq:pnorm}) can be evaluated using the generating function method introducing the new generating function 
\begin{eqnarray}
R(x, y_{1}, y_{2})
%\nonumber\\
=\sum_{\{\nu\}}\frac{x^{\sum_{k}k\nu_{k}}}{\prod_{k}i^{\nu_{k}}\nu_{k}!}\sum_{\{m\}}\left(\prod_{k}y_{1}^{km_{k}}L_{m_{k}}^{(\nu_{k}-m_{k})}(1/k)(-1)^{m_{k}}\right)\sum_{\{r\}}\left(\prod_{k}y_{2}^{kr_{k}}L_{r_{k}}^{(\nu_{k}-r_{k})}(1/k)(-1)^{r_{k}}\right).   
\label{eq:pnorm1}
\end{eqnarray}
The normalization factors of interest Eq. (\ref{eq:pnorm}) can be found as the expansion terms accompanying the products $x^{n}y_{1}^{p}y_{2}^{p}$. 

The generating function can be calculated similarly to Eq. (\ref{eq:Genrecursp1}) evaluating the sums over specific states $k$ as 
\begin{eqnarray}
S_{k}=\exp\left[\frac{y_{1}^{k}+y_{2}^{k}+x^{k}(1-y_{1}^{k})(1-y_{2}^{k})}{k}\right]
\label{eq:pnormderspk}
\end{eqnarray} 
and then taking the product of all these expressions for $k=1, 2...$. Then the generating function can be expressed as 
\begin{eqnarray}
R(x, y_{1}, y_{2})=\frac{(1-xy_{1})(1-xy_{2})}{(1-y_{1})(1-y_{2})(1-x)(1-xy_{1}y_{2})}. 
\label{eq:pnormF}
\end{eqnarray} 
It is convenient to reexpress this generating function as 
\begin{eqnarray}
R(x, y_{1}, y_{2})=\frac{1}{(1-y_{1})(1-y_{2})}+\frac{x}{(1-x)(1-xy_{1}y_{2})}. 
\label{eq:pnormF1}
\end{eqnarray} 
The first term does not contribute to the terms of interest ($x^{n}y_{1}^{p}y_{2}^{p}$ with $n\geq p$) while for the second all expansion coefficients for $n\geq 1$ are equal unity, which proves the normalization of the states $\Psi(p)$ by $1$. 

\section{Solutions for principal quantum numbers $n \leq 5$}

Below we describe the construction of the basis of eigenstates of the anharmonic Hamiltonian $\widehat{V}$ Eq. (\ref{eq:Hfin}) for principal quantum numbers $n \leq 5$ to illustrate the algorithm proposed within the main body of the manuscript. We use the strictly decreasing sequences of quantum numbers $\{ k\}$ satisfying the additional constraint $k_{i-1}-k_{i} \geq k_{i}-k_{i+1}$. One can easily check that all obtained solutions correspond to the wavefunctions normalized by $1$ and orthogonal to each other and all obtained energies are consistent with Eq. 6 in the main text defining energies as a function of generating sequence in the form 
\begin{eqnarray}
\epsilon(\{k\})=-\frac{n(n-1)}{2}+\sum_{i=0}^{p-1}\left[k_{i}(k_{i}-1)-k_{i}k_{i+1}\right].
\label{eq:EigEna}
\end{eqnarray} 
In addition to constructing the eigenstate basis we also discuss the symmetries of the states and their overlap with a single phonon state derived in the main text. 

\subsection{n=1}
\label{sub:n1}

Here the basis consists of only one partition of the number $1$ which is $\nu_{1}=1$. The anharmonic Hamiltonian is represented by a zero diagonal element. The only available sequence in this case is $(1, 0)$ corresponding to all unity solutions $c_{\{1\}}=1$ and zero energy in agreement with Eq. (\ref{eq:EigEna}). 

\subsection{n=2}
\label{sub:n2}

Here the basis consists of two partitions $(\nu_{1}=0, \nu_{2}=1)$ and $(2, 0)$. The operator $\widehat{h}$ Eq. (\ref{eq:OrigPr}) can be represented within this basis as 
\begin{eqnarray}
\widehat{h}_{2} = \begin{pmatrix} 0&1\\ 1&0 \end{pmatrix}.
\label{eq:h2}
\end{eqnarray}
There are two generating sequences including $(2, 0)$ and $(2, 1, 0)$. The first one is represented by all unity eigenvector $\begin{pmatrix} 1\\ 1 \end{pmatrix}$, while the second one should be expressed using the Laguerre polynomials basis set with the second quantum number $m=1$. Since there is only one partition for $m=1$ (see Sec. \ref{sub:n1}) the solution can be expressed in terms of the single Laguerre polynomial $L_{1}^{(\nu_{1}-1)}(1)=\nu_{1}-1=\begin{pmatrix} -1\\ 1 \end{pmatrix}$, while the coefficient with this polynomial has to be one because of the final step $1\rightarrow 0$ corresponding to the all ones solution. The energies of these states are given by $\pm 1$, respectively, in agreement with the analytical prediction of Eq. (\ref{eq:EigEna}). 

Two eigenstates can be expressed through each other using the transformation $\epsilon \rightarrow -\epsilon$, $c_{\{\nu\}}\rightarrow c_{\{\nu\}}\cdot (-1)^{\sum_{i}\nu_{i}}$ described in the main text. Both states contain the single phonon state $(0, 1)$ with the amplitude absolute value equal to $1$ in agreement with the derivation in the main text.

\subsection{n=3}

Here the basis consists of three partitions $(0, 0, 1)$, $(1, 1, 0)$ and $(3, 0, 0)$. The operator $\widehat{h}$ Eq. (\ref{eq:OrigPr}) can be represented within this basis as 
\begin{eqnarray}
\widehat{h}_{3} = \begin{pmatrix} 0&3&0\\ 2&0&1 \\ 0&3&0 \end{pmatrix}
\label{eq:h3}
\end{eqnarray}
There are three generating sequences satisfying the algorithm proposed within the main text including $(3, 0)$, $(3, 1, 0)$ and $(3, 2, 1, 0)$. The first one generates the stationary state given by the all ones eigenvector $\begin{pmatrix} 1\\ 1\\ 1 \end{pmatrix}$, and the second one generates the state given by $L_{1}^{\nu_{1}-1}(1)=\nu_{1}-1=\begin{pmatrix} -1\\ 0\\ 2 \end{pmatrix}$ similarly to Sec. \ref{sub:n2}. The third one should be expressed by the superposition of Laguerre polynomials $L_{2}^{(\nu_{1}-2)}(1)=\nu_{1}(\nu_{1}-1)/2-\nu_{1}+1/2=\begin{pmatrix} 1/2\\ -1/2 \\1/2 \end{pmatrix}$  (partition $(2, 0)$ for $p_{1}=2$) and $L_{1}^{(\nu_{2}-1)}(1/2)=\nu_{2}-1/2=\begin{pmatrix} -1/2\\ 1/2\\-1/2 \end{pmatrix}$ (partition $(0, 1)$ for $p_{1}=2$). The coefficients with these polynomials are determined by the remaining subsequence  $(3, 2, 1, 0)_{-}=(2, 1, 0)$ corresponding to the stationary state $\begin{pmatrix} -1\\ 1 \end{pmatrix}$ (coefficients $\nu_{1}-1$ for the partitions of $p_{1}=2$, see Sec. \ref{sub:n2}) leading to the   final expression for the wavefunction amplitudes 
\begin{eqnarray}
\psi_{(3, 2, 1, 0)}=L_{2}^{(\nu_{1}-2)}(1)-L_{1}^{(\nu_{2}-1)}(1/2)=\begin{pmatrix} 1\\ -1\\ 1 \end{pmatrix}. 
\label{eq:eigvectn3last}
\end{eqnarray} 
The energies of the associated stationary states are given by $\epsilon_{(3, 0)}=3$, $\epsilon_{(3, 1, 0)}=0$, $\epsilon_{(3, 2, 1, 0)}=-3$ in agreement with the analytical prediction of Eq. (\ref{eq:EigEna}). 

Pair of eigenstates described by the sequences $(3, 0)$ and $(3, 2, 1, 0)$ can be expressed through each other using the transformation $\epsilon \rightarrow -\epsilon$, $c_{\{\nu\}}\rightarrow c_{\{\nu\}}\cdot (-1)^{\sum_{i}\nu_{i}}$ described in the main text while the state determined by the sequence $(3, 1, 0)$ transfers to itself. All three states contain the single phonon state $(0, 0, 1)$ with the amplitude absolute value equal to $1$ in agreement with the derivation in the main text.

\subsection{n=4}

Here the basis consists of five partitions including $(0, 0, 0, 1)$, $(0, 2, 0, 0)$, $(1, 0, 1, 0)$,  $(2, 1, 0, 0)$ and $(4, 0, 0, 0)$. The operator $\widehat{h}$ Eq. (\ref{eq:OrigPr}) can be represented within this basis as 
\begin{eqnarray}
\widehat{h}_{4} = \begin{pmatrix} 0&2&4&0&0\\ 4&0&0&2&0 \\ 3&0&0&3&0 \\ 0&1&4&0&1 \\ 0&0&0&6&0 \end{pmatrix}
\label{eq:h4}
\end{eqnarray}
There are five generating sequences including $(4, 0)$, $(4, 1, 0)$, $(4, 2, 0)$, $(4, 2, 1, 0)$ and $(4, 3, 2, 1, 0)$. The first sequence corresponds to all ones eigenstate with energy $\epsilon_{4,0}=6$. The second sequence corresponds to the state 
\begin{eqnarray}
\psi_{(4, 1, 0)}=L_{1}^{(\nu_{1}-1)}(1)=\nu_{1}-1=\begin{pmatrix} -1\\ -1\\ 0 \\ 1 \\3  \end{pmatrix}  
\label{eq:eigvectn4Sec}
\end{eqnarray} 
with energy  $\epsilon_{4,1,0}=2$. The third sequence creates the state corresponding to the symmetric combination of two Laguerre polynomials because the coefficient with polynomials do suppose to be equal one for the terminating step $2\rightarrow 0$. This state can be expressed as 
\begin{eqnarray}
\psi_{(4, 2, 0)}=L_{2}^{(\nu_{1}-2)}(1)+L_{1}^{(\nu_{2}-1)}(1/2)%=
%\nonumber\\
=\nu_{1}(\nu_{1}-1)/2-\nu_{1}+\nu_{2}
=\begin{pmatrix} 0\\ 2\\ -1 \\ 0 \\2  \end{pmatrix}  
\label{eq:eigvectn4Thrd}
\end{eqnarray} 
with energy  $\epsilon_{4,2,0}=0$. The fourth sequence corresponds to anti-symmetric combination of the same functions as in Eq. (\ref{eq:eigvectn4Thrd}) formed similarly to Eq. (\ref{eq:eigvectn3last}) as 
\begin{eqnarray}
\psi_{(4, 2, 1, 0)}=L_{2}^{(\nu_{1}-2)}(1)-L_{1}^{(\nu_{2}-1)}(1/2)%=
%\nonumber\\
=\nu_{1}(\nu_{1}-1)/2-\nu_{1}-\nu_{2}+1=
\begin{pmatrix} 1\\ -1\\ 0 \\ -1 \\3  \end{pmatrix}  
\label{eq:eigvectn4Frth}
\end{eqnarray} 
with energy  $\epsilon_{4,2,1,0}=-2$. The eigenstate determined by the remaining sequence $(4,3,2,1,0)$ can be expressed using the algorithm Eq. (\ref{eq:SolAnz}) and the reduced sequence solution $\psi_{(3, 2, 1, 0)}$ Eq. (\ref{eq:eigvectn3last}) as  
\begin{eqnarray}
\psi_{(4, 3, 2, 1, 0)}=\psi_{(3,2,1,0)}(1) L_{3}^{(\nu_{1}-3)}(1)+\psi_{(3,2,1,0)}(2) L_{1}^{(\nu_{1}-1)}(1)L_{1}^{(\nu_{2}-1)}(1/2)+\psi_{(3,2,1,0)}(3) L_{1}^{(\nu_{3}-1)}(1/3)=
\nonumber\\
=\left[-\frac{1}{6}+\frac{\nu_{1}}{2}-\frac{\nu_{1}(\nu_{1}-1)}{2}+\frac{\nu_{1}(\nu_{1}-1)(\nu_{1}-2)}{6}\right]
-(\nu_{1}-1)(\nu_{2}-1/2)+\nu_{3}-1/3
=\begin{pmatrix} -1\\ 1\\ 1 \\ -1 \\1  \end{pmatrix}. 
\label{eq:eigvectn4Fin}
\end{eqnarray} 
The energy corresponding to this sequence is given by $-6$.

All energies agree with the theory predictions Eq. (\ref{eq:EigEna}).
Pairs of eigenstates described by sequences $(4, 0)$ and $(4, 3, 2, 1, 0)$ and $(4, 1, 0)$, $(4, 2, 1, 0)$ can be expressed through each other using the transformation $\epsilon \rightarrow -\epsilon$, $c_{\{\nu\}}\rightarrow c_{\{\nu\}}\cdot (-1)^{\sum_{i}\nu_{i}}$ described in the main text while the state determined by the sequence $(4, 2, 0)$ transfers to itself. Four states determined by the sequences $(4, 0)$, $(4, 1, 0)$, $(4, 2, 1, 0)$ and $(4, 3, 2, 1, 0)$ contain the single phonon state $(0, 0, 0, 1)$ with the amplitude absolute values equal to $1$ in agreement with the derivation in the main text.

\subsection{n=5}

Here the basis consists of seven partitions $(0, 0, 0, 0, 1)$, $(0, 1, 1, 0, 0)$, $(1, 0, 0, 1, 0)$, $(1, 2, 0, 0, 0)$, $(2, 0, 1, 0, 0)$, $(3, 1, 0, 0, 0)$ and $(5, 0, 0, 0, 0)$. The operator $\widehat{h}$ Eq. (\ref{eq:OrigPr}) can be represented within this basis as 
\begin{eqnarray}
\widehat{h}_{5} = \begin{pmatrix} 0&5&5&0&0&0&0\\ 6&0&0&3&1&0&0 \\4&0&0&2&4&0&0\\0&4&4&0&0&2&0\\0&1&6&0&0&3&0\\
0&0&0&3&6&0&1\\0&0&0&0&0&10&0
\end{pmatrix}
\label{eq:h5}
\end{eqnarray}
The sequences corresponding to the eigenstate basis are given by $(5, 0)$, $(5, 1, 0)$, $(5, 2, 0)$, $(5, 2, 1, 0)$,  $(5, 3, 1, 0)$, $(5, 3, 2, 1, 0)$ and $(5, 4, 3, 2, 1, 0)$. Corresponding eigenstates calculated using the algorithm Eq. (\ref{eq:SolAnz}) similarly to the previously considered cases can be expressed as 
\begin{eqnarray}
\begin{pmatrix} 1\\ 1\\ 1 \\ 1 \\1 \\1 \\1  \end{pmatrix}, ~ \begin{pmatrix} -1\\ -1\\ 0 \\ 0 \\1 \\2 \\4  \end{pmatrix}, ~ \begin{pmatrix} 0\\ 1\\ -1 \\ 1 \\-1 \\1 \\5  \end{pmatrix}, ~ \begin{pmatrix} 1\\ 0\\ 0 \\ -2 \\0 \\0 \\6  \end{pmatrix}, ~ \begin{pmatrix} 0\\ -1\\ 1 \\ 1 \\-1 \\-1 \\5  \end{pmatrix}, ~ \begin{pmatrix} -1\\ 1\\ 0 \\ 0 \\1 \\-2 \\4  \end{pmatrix}, ~ \begin{pmatrix} 1\\ -1\\ -1 \\ 1 \\1 \\-1 \\1  \end{pmatrix}.
\label{eq:n5Ans}
\end{eqnarray}
The corresponding energies of stationary states are given by $10$, $5$, $2$, $0$, $-2$, $-5$, $-10$. 

All energies agree with the theory predictions Eq. (\ref{eq:EigEna}).
Pairs of eigenstates described by sequences $(5, 0)$ and $(5, 4, 3, 2, 1, 0)$, $(5, 1, 0)$ and $(5, 3, 2, 1, 0)$, and $(5, 2, 0)$ and $(5, 3, 1, 0)$   can be expressed through each other using the transformation $\epsilon \rightarrow -\epsilon$, $c_{\{\nu\}}\rightarrow c_{\{\nu\}}\cdot (-1)^{\sum_{i}\nu_{i}}$ described in the main text while the state determined by the sequence $(5, 2, 1, 0)$ transfers to itself. Five states determined by the sequences $(5, 0)$, $(5, 1, 0)$, $(5, 2, 1, 0)$, $(5, 3, 2, 1, 0)$ and $(5, 4, 3, 2, 1, 0)$ contain the single phonon state $(0, 0, 0, 1)$ with the amplitude absolute value equal to $1$ in agreement with the derivation in the main text. 

\section{Matlab programs to calculate Hamiltonian, eigenstates and eigenenergies of the FPU $\alpha$ problem}

Below we describe the supplied Matlab functions targeted to calculate eigenstates and eigenenergies of the problem Eq. (\ref{eq:OrigPr}) using the proposed algorithm Eq. (\ref{eq:wfrecursMatr1}). The verification of the theory using these functions is straightforward. For instance consider the quantum number sequence $(5, ~3, ~1, ~0)$. The eigenfunction  
and eigenenergy corresponding to this sequence can be found numerically using  the command line call ``$[v, E, vn] = EigSt([5, ~ 3, ~1, ~0]);$''. The outcomes include the eigenvector $v$ of the modified problem Eq. (\ref{eq:OrigPr}), the normalized by $1$ eigenvector $vn$ and the eigenenergy $E=-2$ all calculated using the recursive algorithm based on Eq. (\ref{eq:wfrecursMatr1}). The result can be verified generating the system Hamiltonian as ``$y = InitAnhHInfN(5);$''. Then the standard Hamiltonian ``$\widehat{H}$'' can be found using ``$H=y.H$''. The operation ``$H$*$vn$-$E$*$vn$'' should return a vector of zeros (with the appropriate accuracy which is $10^{-15}$ in my computer; the equation is typed in the way that it can be copied and pasted directly to the Matlab command line at least in my computer). The modified representation of the Hamiltonian can be found using the command ``$y1 = InitAnhHInfNMod(5);$''. Then the operation ``$y1.H$*$v$-$E$*$v$'' should lead to a vector of zeros as well. The reader is strongly encouraged to try these codes for different quantum number sequences to examine the validity of the proposed solution.  

Before running the  commands you need to copy all supplied Matlab files to the same folder and change the current folder used by Matlab to that specific folder. 

\subsection{Function $y = FindResModes(M, S, N)$} 

This function finds all integer partitions of a number $S$ made using $N-M+1$  integer numbers $M, M+1,... N$. It is targeted to find  the basis set for the problem of interest expressed in terms of phonon population numbers.  The call ``$y = FindResModes(1, N, N)$'' returns all integer partitions of the number $N$. Partitions are expressed as rows of the matrix representing the answer. 

The number $S$ must be less or equal $N$ and  the parameter $M$ must be greater or equal to $1$ and less or equal to both $N$ and $S$. This function should work fine until $N \leq 60$. 

This function is independent of other author's programs.

% Hamiltonian 
\subsection{Function $y = InitAnhHInfN(N)$} 

This function generates resonant Hamiltonian of the third order anharmonic interactions within the long-wavelength and large size limits for the principal quantum number $N$.  

The outcome $y.H$ returns the Hamiltonian matrix, the other outcome $y.Hsp$ returns the same matrix in 
the sparse matrix Matlab form (type ``$y = InitAnhHInfN(5);$ $y.H$'' in the Matlab command  line to generate the Hamiltonian for $N=5$).

The number $N$ must be an integer number greater than $1$ and less or equal to $30$;  for $N>30$ the memory can be insufficient 
for the resulting Hamiltonian matrix. Then the only sparse matrix outcome can be used.

This function depends on the function "FindResModes" generating the basis set of partitions.

\subsection{Function $y = InitAnhHInfNMod(N)$} 

 This function generates modified Hamiltonian $\widehat{h}$ (Eq. (\ref{eq:OrigPr})) of the third order anharmonic interactions within the long-wavelength and large size limits for the principal quantum number $N$.  

The outcomes $y.H$ returns the Hamiltonian matrix, the other outcome $y.Hsp$ returns the same matrix in  the sparse matrix Matlab form  (type ``$y = InitAnhHInfNMod(5);$ $y.H$'' in the Matlab command  line to generate the modified Hamiltonian for $N=5$)

 The number $N$ must be an integer number greater than $1$ and less or equal to $30$;  for $N>30$ the memory can be insufficient 
for the resulting Hamiltonian matrix. Then the only sparse matrix outcome can be used.
     
This function depends on the function "FindResModes" generating the basis set of partitions.

\subsection{Function $[y, E, yn] = EigSt(B)$} 

 This function generates the eigenstate and the related eigenenergy using the sequence of quantum numbers $B$ determining this state. 

The vector $y$ returns the eigenstate amplitudes in the modified population number representation and the number $E$ returns the eigenstate energy,  while $yn$ returns the wavefunction amplitudes normalized by $1$ in the true population number representation.

The set  $B$ must be a strictly decreasing set of integer numbers beginning with the principal quantum number $n$ and ending with $0$ (For example eigenstate and eigenenergy corresponding to the set $(5, 3, 2, 1, 0)$ can be found typing ``$[y, E] = EigSt([5, ~3, ~2, ~1, ~0])$'').  
 
This function uses the functions ``FindResModes" to generate partitions, ``BasFun" to make the wavefunction expansion over the basis of the polynomial products and "NormFact" to switch to the standard basis.  
 
\subsection{Function $y = CollectsEigsPartit(N)$} 
 
 This function is expected to generate the full set of eigenstates and the related 
eigenenergies using the special set of sequences of strictly decreasing quantum numbers  with the constraint that the array of differences for any sequences is non-decreasing (the set $(5, 3, 2, 1, 0)$ is acceptable, while the set $(5, 3, 0)$ is not because $5-3 < 3-0$). 

The matrix $y.V$ returns eigenvectors in a modified population number representation as columns, the row $y.Etst$ returns eigenenergies corresponding to eigenvectors and the matrix $y.Comb$ returns the matrix of sequences used to generate the eigenstates. 

The completeness of the basis set can be tested calculating the rank of the matrix of eigenvectors ``$A=rank(y.V);$". The result can then be compared with the size of the matrix of eigenvectors that can be determined using the command ``$S=rank(y.V);$". According to our consideration the rank of the matrix is identical to both sizes of that matrix at least up to $N=25$. Unfortunately, we cannot prove the completeness of the basis set analytically for arbitrarily $N$. 

$N$ must be a positive integer number. 

This function uses the functions ``FindResModes" to generate partitions and ``EigSt" to find eigenstates and energies for each sequence.  

\subsection{Function $y = NormFact(P)$} 

This function generates the normalization factors for conversion between actual (normalized by $1$) and modified basis sets. 

 The vector $y$ returns the column of normalization factors for each row of the input matrix $P$. 

Matrix $P$ is a matrix of sets (rows) of population numbers representing 
some basis states. All numbers should be integer and nonnegative.  
  
   This function is independent of other author's programs.

\subsection{Function $y = BasFun(Part, Bas)$} 

This function calculates the basis functions as a products of Laguerre polynomials Eq. (\ref{eq:LaguerreProdApp}).
   The sequence $Part$ represents the sequence of population numbers (partition) and 
 the matrix $Bas$ represents the sets of orders of corresponding Laguerre
polynomials which can be expressed as the set of partitions. 
 
 Vector $y$ returns the vector of the products of polynomials of the specific 
 population number set $Part$ for each  row of the sequence $Bas$. 
 For instance if the function arguments are the matrix $Bas=\begin{pmatrix} 1 & 1 & 0 \\ 0 & 0 & 1 \end{pmatrix}$ and the partition $Part=(5, 2, 1)$ then 
the outcome vector will be  $y=\begin{pmatrix} L_{1}^{(4)}(1)L_{1}^{(1)}(1/2)L_{0}^{(1)}(1/3)\\ L_{0}^{(5)}(1)L_{0}^{(2)}(1/2)L_{1}^{(0)}(1/3) \end{pmatrix}$. 
 
   All sequences should contain only integer nonnegative numbers. 
  
   This function uses the function "Coeft" evaluating Laguerre polynomials.  

\subsection{Function $y = Coeft(r, k, n)$} 

This function calculates the basis function for the individual state $k$ with the population number $n$ expressed in  terms of associated Laguerre polynomials \cite{Laguerre} as  $L_r^{(n-r)}(1/k)$. 

Inputs $r$ and $n$ must be nonnegative integer numbers.

%\begin{thebibliography}{100}

%\bibitem{FPUclassic}
%E. Fermi, J. Pasta, S. Ulam,  Studies of nonlinear problems. No. LA 1940. I, Los Alamos Scientific Laboratory Report %No. LA-1940 (Los Alamos Scientific Laboratory, Los Alamos, NM, (1955)). 

%\bibitem{Laguerre}
%The generalized Laguerre polynomials can be defined as $L_{m}^{(n-m)}(x)=\sum_{p=0}^{min(m, n)}\frac{(-1)^{m-p} x^{m-p}n!}{p!(n-p)!(m-p)!}$ for integer indices $m$ and $n$, see  %\cite{Laguerre1,Gradshtein}.  %We will use two %properties of Laguerre polynomials  $aL_{a}^{(b)}(x)=(a+b)L_{a-1}^{(b)}(x)-xL_{a-1}^{(b+1)}(x)$ and $L_{a}^{(b)}%(x)=L_{a}^{(b+1)}(x)-L_{a-1}^{(b+1)}(x)$ which can be verified substituting their definitions, see  
%
%\bibitem{Laguerre1}
%M. Abramowitz, I. A. Stegun,  eds., Handbook of Mathematical Functions with Formulas, Graphs, and Mathematical %Tables, New York: Dover, p. 773 (1965).

%\bibitem{ProofDescr}
%The proof of the normalization of the wavefunction described by the sequence $(n, k, k-1, k-2,...1)$ by $1$ can be performed in two steps. First one consider the sequence $(k, k-1, k-2,...1, 0)$ and show that the related wavefunction can be chosen in the form $c_{\{k\}}=(-1)^{\sum_{i}k_{i}}$ for the given partition of the number $k$ which is normalized by $1$ since the absolute values of the wavefunction amplitudes are identical to those of the normalized by $1$ state $(k, 0)$. The proof of the normalization by one for the state $(n, k, k-1, k-2,...1)$ can be made then using its wavefunction definition Eq. (\ref{eq:wfrecurs1}). 

%\end{thebibliography}

%\end{document}
\end{widetext}


\begin{thebibliography}{99}

\bibitem{AbeScience1}  
A. Nitzan and M. A. Ratner, 
%{\em Electron transport in molecular wire functions: Models and Mechanisms}, 
Science {\bf 300}, 1384 (2003). 




\bibitem{AbeScience2}
A. Nitzan, 
%{\em Molecules take the heat},   
Science {\bf 317}, 759 (2007). 

\bibitem{RevModPh}
N. Li, J. Ren, L. Wang, G. Zhang, P. Hänggi, and B. Li,
%{\em Colloquium: Phononics: Manipulating heat flow with electronic analogs and beyond}, 
Rev. Mod. Phys. {\bf 84}, 1045 (2012).

\bibitem{GruebeleQC1}
D. Weidinger, and M. Gruebele, 
%{\em Quantum computation with vibrationally excited polyatomic molecules: effects of rotation, level structure, and field gradients},
Mol. Phys. {\bf 105}, 1999 (2007).

\bibitem{GruebeleQC2}
D. Shyshlov, E. Berrios, M. Gruebele, and D. Babikov,
%{\em On readout of vibrational qubits using quantum beats}, 
J. Chem. Phys. {\bf 141}, 224306 (2014). 

\bibitem{Shapiro}
H. Li, T. Kottos, and B. Shapiro, 
Phys. Rev. E {\bf 91}, 042125 (2015).  

\bibitem{DlottScience}
Z. Wang, J. A. Carter,  A. Lagutchev, Y. K. Koh, N.-H. Seong, D. C. Cahill, D. D. Dlott, 
%{\em Ultrafast Flash Thermal Conductance of Molecular Chains}, 
Science {\bf 317}, 787 (2007). 

%\bibitem{PolyethilenePRL}
%A. Henry, and G. Chen,  
%{\em High Thermal Conductivity of Single Polyethylene Chains Using Molecular Dynamics Simulations}, 
%Phys. Rev. Lett. {\bf 101}, 235502 (2008). 



\bibitem{NitzanSegal1}
D. Segal, A. Nitzan, and P. Hänggi, 
%{\em Thermal conductance through molecular wires}, 
J. Chem. Phys. {\bf 119}, 6840 (2003). 

\bibitem{Alkane3}
T. Meier, F. Menges, P. Nirmalraj, H. Holscher, H. Riel, and B. Gotsmann, 
%{\em Length-Dependent Thermal Transport along Molecular Chains}, 
Phys. Rev. Lett. {\bf 113}, 060801  (2014).

\bibitem{IgorRecent}
N. I. Rubtsova, C. M. Nyby, H. Zhang, B. Zhang, X. Zhou, J. Jayawickramarajah, A. L. Burin. and I. V. Rubtsov, 
%{\em Room-temperature ballistic energy transport in molecules with repeating units}, 
J. Chem. Phys. {\bf 142}, 212412 (2015). 

%{Turbulence,Leibovitz2008,Shepel2009,Flach2009,Leibowitz2011,ab2015}




\bibitem{DharRev}
%Heat transport in low-dimensional systems
A. Dhar, Adv. Phys. {\bf 57}, 457 (2008). 

\bibitem{abPhR}
I. Y. Polishchuk,  L. A. Maksimov, A. L. Burin, Phys. Rep. {\bf 288}, 205 (1997). 


%\bibitem{RevRecent1}
%A.J. Lichtenberg, R. Livi, M. Pettini, and S. Ruffo,  
%Dynamics of Oscillator Chains, 
%Lect. Notes Phys. {\bf 728}, 21 (2008).

\bibitem{Israilev}
G. P. Berman, and F. M. Izrailev, 
% ??The Fermi-Pasta-Ulam problem: Fifty years of progress,?? 
Chaos {\bf 15}, 015104 (2005).

\bibitem{RevRecent}
G. Benettin, H. Christodoulidi, A. Ponno, J. Stat. Phys. {\bf 152}, 195 (2013); A.J. Lichtenberg, R. Livi, M. Pettini, and S. Ruffo, Lect. Notes Phys. {\bf 728}, 21 (2008).
%The Fermi-Pasta-Ulam problem and its underlying integrable dynamics. J Stat Phys 152(2):195 ?212



\bibitem{FPUclassic}
E. Fermi, J.R. Pasta and S. Ulam, in Collected papers of Enrico Fermi, E. Segre (ed.). University of Chicago Press, Chicago, {\bf 2}, 978 (1965).
%E. Fermi, J. Pasta, S. Ulam,  Studies of nonlinear problems. No. LA 1940. I, Los Alamos Scientific Laboratory Report No. LA-1940 (Los Alamos Scientific Laboratory, Los Alamos, NM, (1955)). 


%\bibitem{AlkaneSpectrum} 
%L. Piseri, G. Zerbi, %{\em Dispersion Curves and Frequency Distribution of Polymers: Single Chain Model}, 
%J. Chem. Phys. {\bf 48}, 3561 (1968).% 3561-3571. 

%\bibitem{SolitonCl}
%N. J. Zabusky, M. D. Kruskal,  Interaction of Solitons in a Collisionless Plasma and the Recurrence of Initial State, (Princeton University Plasma Physics Laboratory, Princeton, 1965).

\bibitem{Palais}
R. S. Palais, Bull. (New Series) of Amer. Math. Soc. {\bf 34}, 339 (1997).  


%\bibitem{Zakharov}
%V. E. Zakharov, E. I. Schulman, Integrability of nonlinear systems and perturbation theory. What Is Integrability? ed V. E. Zakharov (Springer, Berlin), 185 (1991).

\bibitem{Henrici}
A. Henrici, T. Kappeler, Commun. Math. Phys. {\bf 278}, 145 (2008). 

\bibitem{Doron}
I. Tikhonenkov, A. Vardi, J. R. Anglin, and D. Cohen, %{\em Minimal Fokker-Planck Theory for the Thermalization of Mesoscopic Subsystems}, 
Phys. Rev. Lett 110 (2013), 050401. 

\bibitem{PNASRecent}
M. Onoratoa, L. Vozellaa, D. Promentb, and Y. V. Lvov, PNAS {\bf 112}, 4208 (2015).  

\bibitem{Flach}
S. Flach, A. V. Gorbach, Phys. Rep. {\bf 467}, 1 (2008). % 1?116

\bibitem{SI}
Supplementary materials contain Matlab codes and mathematical derivations of some statements made within the main body of the manuscript.  

\bibitem{TextPh}
J.A. Reissland, The Physics of Phonons. (Wiley, Toronto, 1972). 



\bibitem{PartitionTheory}
G. E. Andrews, The Theory of Partitions (1976), Cambridge University Press. ISBN 0-521-63766-X. 


\bibitem{Laguerre}
The associated Laguerre polynomials can be defined as $L_{a}^{(b)}(x)=\sum_{p=0}^{a}\frac{(-1)^{a-p} x^{a-p}}{p!(a-p)!}\frac{(a+b)(a+b-1)...(a+b-p+1)}{p!}$ for integer indices $a$ and $b$, %\cite{Laguerre}.  %We will use two properties of Laguerre polynomials  $aL_{a}^{(b)}(x)=(a+b)L_{a-1}^{(b)}(x)-xL_{a-1}^{(b+1)}(x)$ and $L_{a}^{(b)}(x)=L_{a}^{(b+1)}(x)-L_{a-1}^{(b+1)}(x)$ which can be verified substituting their definitions. 
%
%\bibitem{Laguerre1}
see M. Abramowitz, I. A. Stegun,  eds., Handbook of Mathematical Functions with Formulas, Graphs, and Mathematical Tables, New York: Dover, p. 773 (1965).

\bibitem{Lukin2}
N. Y. Yao, C. R. Laumann, S. Gopalakrishnan, M. Knap, M. M\"uller, E. A. Demler, M. D. Lukin, 
%Many-body Localization in Dipolar Systems, 
Phys. Rev. Lett. {\bf 113}, 243002 (2014).

%\bibitem{abChaos}
%A. L. Burin, 
%Many-body delocalization in a strongly disordered system with long-range interactions: Finite-size scaling, 
%Phys. Rev. B {\bf 91}, 094202 (2015). 

\bibitem{LeitnerArnoldDiff} 
D. M. Leitner, P. G. Wolynes, %{\em Quantization of the Stochastic Pump Model of Arnold Diffusion}, 
Phys. Rev. Lett. {\bf 79}, 55 (1997).

\bibitem{LeibovitzLocalNonlin}
A. Dhar, J. L. Lebowitz, %Effect of phonon-phonon interactions on localization, 
Phys. Rev. Lett. {\bf 100}, 134301 (2008).  


\bibitem{David1}
D. M. Leitner, 
%Vibrational energy transfer and heat conduction in a one-dimensional glass
Phys. Rev. B {\bf 64}, 094201 (2001). 

\bibitem{Gradshtein}
I.S. Gradshteyn and I.M. Ryzhik, Eds. A. Jeffrey, D. Zwillinger, Table of Integrals, Series, and Products, 7$^{th}$ editions, Elsevier, p. 1002, Eq. 8.975, 2.  

\end{thebibliography}
\end{document}